\documentclass[a4paper,fleqn,usenatbib,useAMS]{mnras}

\usepackage{newtxtext,newtxmath}

\usepackage[T1]{fontenc}
\usepackage{ae,aecompl}
\usepackage{lineno}
\usepackage{color}
\usepackage{graphicx}	
\usepackage{amsmath}	
\usepackage{amssymb}	
\usepackage{multicol}        
\usepackage{bm}		
\usepackage{multirow}
\usepackage{enumitem}
\usepackage{xr}
\normalsize

\title[Hierarchically Modelling the Ages of WDs]{A Hierarchical Model for the Ages of Galactic Halo White Dwarfs}

\author[Si, van Dyk, von Hippel et al.]{
Shijing Si,$^{1}$\thanks{Contact e-mail: \href{mailto:ss2913@ic.ac.uk}{ss2913@ic.ac.uk}}
David A. van Dyk,$^{1}$
Ted von Hippel,$^{2,3}$
Elliot Robinson,$^{4}$
\newauthor{Aaron Webster,$^{2}$ David Stenning$^{5}$}
\\
$^{1}$Statistics Section, Department of Mathematics, Imperial College London, London SW7 2AZ, UK\\
$^{2}$Center for Space and Atmospheric Research, Embry-Riddle Aeronautical Univeristy, Daytona Beach, USA\\
$^{3}$Max Planck Institute for Astronomy, K{\"o}nigstuhl 17, 69117 Heidelberg, Germany\\
$^{4}$Argiope Technical Solutions, Florida, USA\\
$^{5}$Institut d'astrophysique de Paris, Paris, France
}

\date{Accepted 2017 March 24. Received 2017 March 24; in original form 2016 August 12}

\pubyear{2017}

\begin{document}
\label{firstpage}
\pagerange{\pageref{firstpage}--\pageref{lastpage}}
\maketitle

\begin{abstract}
In astrophysics, we often aim to estimate one or more parameters for each member object
in a population and study the distribution of the fitted parameters across the population. In this
paper, we develop novel methods that allow us to take advantage of existing software
designed for such case-by-case analyses to simultaneously
fit parameters of both the
individual objects and the parameters that quantify their distribution across the population. 
Our methods are based on Bayesian hierarchical modelling which is known to produce
parameter estimators for the individual objects that are on average closer to their
true values than estimators based on case-by-case analyses. We verify this in the context of
estimating ages of Galactic halo white dwarfs (WDs) via a series of simulation studies. Finally,
 we deploy our new techniques on optical and near-infrared photometry of
ten candidate halo WDs to obtain estimates of their ages
along with an estimate of the mean age of Galactic halo WDs of $12.11_{-0.86}^{+0.85}$ Gyr. 
Although this sample is small, our technique lays the ground work for large-scale
studies using data from the Gaia mission.

\end{abstract}




\section{Introduction}

\subsection{White Dwarfs and The Galactic Halo Age}
In the astrophysical hierarchical structure formation model, the present Galactic stellar
halo is the remnant of mergers of multiple smaller galaxies \citep[e.g.][]{freeman2002new, tumlinson2010cosmological, scannapieco2011formation}, most of
which presumably formed some stars prior to merging, and some of which may
have experienced, triggered, or enhanced star formation during the merging
process.  The age distribution of Galactic halo stars encodes this process.
Any perceptible age spread for the halo thus provides information on this
complex star formation history.

At present, we understand the Galactic stellar halo largely through the
properties of its globular clusters.  These star clusters are typically
grouped into a few categories: i) those with thick disk kinematics and
abundances, ii) those with classical halo kinematics and abundances, iii) the
most distant population that is a few Gyr younger than the classical halo
population, and iv) a few globular clusters such as M54 that are ascribed to
known, merging systems, in this case the Sagittarius dwarf galaxy 
\citep[see][]{forbes2010accreted, pawlowski2012vpos, leaman2013bifurcated}.
Globular clusters in category two
appear consistent with the simple collapse picture of \citet{eggen1962evidence},
 yet those of categories three and four argue for a more
complex precursor plus merging picture.  The newly appreciated complexity
of multiple populations in many or perhaps all globular clusters \citep{gratton2004abundance}
 adds richness to this story, and may eventually
help us better understand the earliest star formation environments.

Despite the tremendous amount we have learned from globular clusters, they
are unlikely to elucidate the full star formation history of the Galactic
halo because today's globular clusters represent a $\sim1$\% minority of
halo stars.  Without studying the age distribution of halo field stars, we
do not know whether globular cluster ages are representative of the entire
halo population.  We do know that globular clusters span a narrower range
in abundances than field halo stars \citep[see][]{roederer2010characterizing, yamada2013stellar},
 so there is every reason to be suspicious that there is more to
the story than globular clusters can themselves provide.

In order to determine the age distribution of the Galactic halo, we need to
supplement the globular cluster-based story with ages for individual halo
stars.  This is not practical for the majority of main sequence or red
giant stars because of well-known degeneracies in their observable
properties as a function of age.  Gyrochronology \citep[see][]{barnes2010simple, soderblom2010ages}
does hold some hope for determining the ages of individual
stars, but this is unlikely to provide precise ages for very old stars even
after the technique sees considerably more development.  Our best current
hope for deriving the Galactic halo distribution is to determine the ages
of halo field WDs.

WDs have the advantages that they are the evolutionary end-state
for the vast majority of stars and their physics is relatively well
understood \citep{fontaine2001potential}.  A WD's surface
temperature, along with its mass and atmospheric type, is intimately
coupled to its cooling age, i.e., how long it has been a WD.  The mass of a
WD, along with an assumed initial-final mass relation (IFMR), provides the
initial main sequence mass of the star, which along with theoretical
models, provides the lifetime of the precursor star.  Pulling all of this
information together provides the total age for the WD.  The weakest link
in this chain is typically the IFMR.  Yet fortunately the uncertainty in
the IFMR often has little effect on the \textit{relative} ages of WDs, and thus the
precision of any derived age distribution.  Additionally, among the higher
mass WDs, the uncertainty in the precursor ages can be reduced to a level
where the IFMR uncertainties do not dominate uncertainties in the absolute
WD ages. 

While WDs provide all of these advantages for understanding stellar ages,
the oldest are very faint, and thus few are known, with fewer still known
with the kinematics of the Galactic halo.  The paucity of data for these
important objects will shortly become a bounty when Gaia both finds
currently unknown WDs with halo kinematics and provides highly accurate and
precise trigonometric parallaxes, which constrain WD surface areas and thus
masses.  The number of cool, halo WDs is uncertain by a factor of perhaps
five, and depending on the Galaxy model employed, \citet{carrasco2014gaia}
calculate that Gaia will derive parallaxes for $\sim60$ or $\sim350$ single halo
WDs with $T_{\rm eff} \leq5000$.  Gaia will measure parallaxes for more
than 200,000 WDs with thick disk and disk kinematics.

We have developed a Bayesian statistical technique to derive the ages of
individual WDs \citep{van2009, o2013} and intend to apply
this to each WD for which Gaia obtains excellent parallaxes.  Yet the
number of halo WDs for which we can derive high-quality ages
may still be modest, particularly because we also
require accurate optical and near-IR photometry.  Because of the importance
of the age distribution among halo stars, we have developed a hierarchical
modelling technique to pool halo WDs and derive the posterior
distributions of their ages.  

\subsection{Statistical Analysis of a Population of Objects}
In statistics, 
hierarchical models are viewed as the most efficient and principled technique
to estimate the parameters of each object in a population \citep[e.g.,][]{gelman2006}.
 In astrophysics, we often aim to estimate a set of parameters for each of a number of objects
in a population, which motivates the application of hierarchical models.
Noticeably, these models have gained popularity in astronomy mainly for two reasons. 
Firstly, they provide an approach to
combining multiple sources of information.
For instance, \citet{mandel2009type} employed Bayesian hierarchical models
to analyse the properties of Type Ia supernova light curves by using data from Peters Automated 
InfraRed Imaging TELescope and the literature. 
Secondly, they generally produce estimates with less uncertainty. 
By combining information from supernovae Ia lightcurves,
\citet{march2014comparison} and \citet{shariff2016standardizing} illustrate
 how a hierarchical model can improve 
the estimates of cosmological parameters.
Similarly, \citet{licquia2015improved} obtained improved estimates
 of several global properties of the Milky Way by using
a hierarchical model to combine previous measurements from the literature.

However, fully modelling a population of objects within a
hierarchical model requires substantial computational investment and often specialised computer code,
especially for complicated problems.
In this study, we develop novel methods to conveniently obtain the improved estimates available under
a hierarchical model. While taking advantage of the existing code for case-by-case analyses,
our methods simultaneously estimate parameters of the individual objects and parameters that
describe their population distribution. 
Our methods are based on Bayesian hierarchical modelling which are known to produce
estimators  of parameters of the individual objects that are on average closer to their
true values than estimators based on case-by-case analyses.

 There are many possible applications of hierarchical models in astrophysics. 
 In this article we focus on the analysis of a sample of
candidate halo WDs. We perform a
simulation study to illustrate the advantage of our approach over the commonly used 
case-by-case analysis in this setting. We find that approximately two thirds of 
the estimated WD ages are closer to
their true age under the hierarchical model. Using optical and near-infrared photometry of
ten candidates halo WDs, we simultaneously estimate their individual ages and
the mean age of these halo WDs; the latter which we estimate as $12.11_{-0.86}^{+0.85}$ Gyr. 
Another application to the distance modulus of the Large Magellanic Cloud (LMC) is included 
in Appendix \ref{sec:lmc} as a pedagogical illustration of our methods in a simpler setting. 

One of the primary benefits of 
 our approach is that it takes advantages of the existing code which
fits one object at a time. We only need to write wrapper code that calls the existing programs, 
see \citet{si2017} for more details.

 This
saves substantial human capital that might otherwise be devoted to developing and coding a complex
new algorithm.
The power of this approach can be conceived of as coming from i) an informative assumption, which
is that all the objects belong to a population with a particular distribution of the parameters of the objects
across the population, and ii) that
it otherwise is difficult to come up with a technique that can combine the individual results
when they may have asymmetric posterior density functions. 

The remainder of this article is organised into five sections.
We introduce hierarchical modelling and its statistical inference methods in Section 2.
We present methods for case-by-case and hierarchical analyses of the ages of a group
of WDs in Section 3. In Section 4, we use a simulation study to verify 
the advantages of the hierarchical approach.
In Section 5 we apply both the case-by-case and our hierarchical model to ten Galactic halo WDs,
and then interpret the Galactic halo age in the context of known Milky Way ages.
Section 6 summarises the proposed methodology and our results.
 In Appendix \ref{sec:shrinkage}, we describe the statistical background of
hierarchical models and explain why they tend to provide better estimates.
We illustrate the application of  
hierarchical models and their advantageous statistical properties
via the LMC example in Appendix \ref{sec:lmc}. In Appendices \ref{sec:fb}
and \ref{sec:mcem} outline the computational algorithms we use to efficiently fit   
the hierarchical models.
  
\section{Hierarchical Modelling}\label{sec:hier0}

Suppose we observe a sample of objects from a population of astronomical sources,
 for example, the photometry of
 $10$ WDs from the Galactic Halo, and
 we wish to estimate a particular parameter or a
set of parameters for each object. We refer to these as the {\it object-level parameters}.
 By virtue of the population, there is a distribution of these parameters
across the population of objects. This distribution can be described by
another set of parameters that we refer to as the {\it population-level parameters}.
Often we aim to estimate both the
object-level parameters and population-level parameters. As we shall see, however,
even if we are only interested in the object-level parameters, they can be better estimated
if we also consider their population distribution.

Hierarchical models \citep[e.g.,][]{gelman2014}, also called random effect models,
 can be used to combine data from multiple objects in a single 
 coherent statistical analysis.
Potentially this can lead to a more comprehensive
understanding of the overall population of objects.
 Hierarchical models are widely used in many fields, spanning the
medical, biological, social, and physical sciences. Because they 
leverage a more comprehensive data set when fitting
the object-level parameters, they tend to result in estimators that on average exhibit smaller errors
\citep[e.g.,][]{james1961,efron1972,carlin2000,morris2012shrinkage}. 
Because a property of these estimators is that they are ``shrunk'' toward a common central value
 relative to those derived from the corresponding case-by-case analyses, they are often called \textit{
 shrinkage estimators}.
More details about shrinkage estimators appear in Appendix \ref{sec:shrinkage}.

A concise hierarchical model is 
 \begin{align}
{Y}_{i}|\theta_{i}&\sim{N}(\theta_{i},\sigma),~i=1,2,\cdots,n,\label{eq:nhm1}\\
\theta_{i}&\sim{N}(\gamma,\tau),\label{eq:nhm2}
\end{align}
where $\bm{Y}=(Y_{1}, \cdots, Y_{n})$ are observations, $\sigma$ is the standard error of the
observations,
$\bm{\theta}=(\theta_{1},\cdots, \theta_{n})$ are objective-level parameters of interest,
$\gamma$ and $\tau$ are the unknown population-level mean and standard deviation parameters
of a Gaussian distribution\footnote{In this paper we parameterise univariate Gaussian distributions in terms of
their means and standard deviations. Generally, we write $Y|\theta\sim{N}(\mu, \sigma)$
to indicate that given $\theta$, $Y$ follows a Gaussian (or Normal) distribution with mean $\mu$ and
standard deviation $\sigma$.
}. 

Bayesian statistical methods use the conditional probability distribution of the unknown
parameters given the observed data to represent uncertainty and generate parameter estimates
and error bars. This conditional probability distribution is called the \textit{posterior distribution}
and in our notation written as $p(\gamma, \tau, \bm{\theta}| \bm{Y})$. To derive the posterior
distribution via Bayes theorem requires us to specify a \textit{prior distribution} which summarises
our knowledge about likely values of the unknown parameters having seen the data, see
\citet{van2001analysis} and \citet{park2008searching} for applications of Bayesian methods in
the context of astrophysical analyses. Our prior distribution on $\bm{\theta}$ is given in Eq. \ref{eq:nhm2}
and we choose the non-informative prior distribution $p(\gamma, \tau)\propto1$ for $\gamma$
and $\tau$, which is a standard choice in this setting \citep{gelman2006prior}.
Two commonly used Bayesian methods to fit the hierarchical model in Eq. \ref{eq:nhm1}--\ref{eq:nhm2} are
the fully Bayesian (FB) and the empirical Bayes (EB) methods.  

FB \citep[e.g.,][]{gelman2014} fits all of the unknown parameters via their joint posterior distribution
\begin{equation}\label{eq:fb_ex}
{p}(\gamma, \tau, \bm{\theta}|\bm{Y})\propto{p}(\gamma, \tau)\prod_{i=1}^{n}{p}(\theta_{i}|\gamma, \tau)\prod_{i=1}^{n}{p}(Y_{i}|\theta_{i}).
\end{equation}
Generally, we employ Markov chain
Monte Carlo (MCMC) algorithms to obtain
a sample from the posterior distribution, $p(\gamma, \tau, \bm{\theta}|\bm{Y})$.
The MCMC sample can be used to i) generate parameter estimates, e.g., by averaging 
the sampled parameter values, ii) generate error bars, e.g., by computing percentiles of the
sampled parameter values, and iii) represent uncertainty, e.g., by plotting histograms
or scatter plots of the sampled values.
For intricate hierarchical models, however it may be computationally challenging to obtain a
reasonable MCMC sample.

 EB \citep[e.g.][]{morris1983,casella1985introduction,efron1996} uses the data to first
 fit the parameters of the prior distribution in
 Eq. \ref{eq:nhm2} and then given these fitted parameters
 infer parameters in Eq. \ref{eq:nhm1} in the standard Bayesian way. Specifically,
 $\gamma$ and $\tau$ are first estimated as $\hat{\gamma}$ and then $\hat{\tau}$ and the prior distribution
 $\theta_{i}\sim{N}(\hat{\gamma}, \hat{\tau})$ is used in a Bayesian analysis to estimate the $\theta_{i}$. 
 Thus, EB proceeds in two steps.
 \begin{description}
 \item[\textbf{Step 1}] Find the maximum a posterior (MAP) estimates of $\gamma$ and $\tau$ by maximising
 their joint posterior distribution, i.e,
 \begin{equation}\label{eq:map1}
 (\hat{\gamma}, \hat{\tau})=\arg\max_{\gamma, \tau}{p}(\gamma,\tau|\bm{Y})=\arg\max_{\gamma, \tau}\int{p}(\gamma,\tau, \bm{\theta}|\bm{Y}){d\bm{\theta}}.
 \end{equation}
\item[\textbf{Step 2}] Use ${N}(\hat{\gamma}, \hat{\tau})$ as the prior distribution for
$\theta_{i}, i=1, \cdots, n$ and estimate $\theta_{i}$ in the standard Bayesian way, i.e.,
\begin{equation}\label{eq:eb_ex2}
 {p}(\theta_{i}|Y_{i}, \hat{\gamma}, \hat{\tau})\propto{p}(Y_{i}|\theta_{i}){p}(\theta_{i}|\hat{\gamma}, \hat{\tau}).
 \end{equation}
\end{description}

When applying the EB to fit a hierarchical model, it is possible that the estimate of
the standard deviation $\tau$ is equal to $0$,
which leads to $\theta_{1}=\cdots=\theta_{n}=\hat{\gamma}$. This is generally not a desirable result.
We can avoid $\hat{\tau}=0$ by using the transformations $\xi=\log\tau$ or $\delta=1/\tau$
\citep[e.g.,][]{park2008searching, gelman2014}.
We refer to EB implemented with these transformations as EB-log and EB-inv, respectively.
\textbf{Step 2} of EB-log and EB-inv remains exactly the same as that of EB, but \textbf{Step 1}
 changes. Specifically, Step 1 of EB-log is 
   \begin{description}
 \item[\textbf{Step 1}] Find the MAP estimates of $\gamma$ and $\xi$ by maximising
 their joint posterior distribution, i.e.,
 \begin{equation*}
 (\hat{\gamma}, \hat{\xi})=\arg\max_{\gamma, \xi}{p}(\gamma,\exp(\xi)|\bm{Y})\exp(\xi)
 \end{equation*}
 and setting $\hat{\tau}=\exp(\hat{\xi})$, where $p(\cdot|\bm{Y})$ is the posterior distribution
 of $\gamma$ and $\tau$. Thus, 
  \begin{equation}\label{eq:map2}
 (\hat{\gamma}, \hat{\tau})=\arg\max_{\gamma, \tau}{p}(\gamma,\tau|\bm{Y})\tau.
 \end{equation}
 Comparing Eq. \ref{eq:map2} with Eq. \ref{eq:map1}, the added $\tau$ in Eq. \ref{eq:map2}
 prevents $\tau$ from being zero. The Step 1 of EB-inv proceeds similarly.
 \end{description}

\section{Analyses for Field Halo White Dwarfs}\label{sec:estimation}

Our model is based on obtaining photometric magnitudes for
 $n$ WDs from the Galactic halo. We denote the $l$-dimensional observed photometric
 magnitudes for the $i$-th WD by $\bm{X}_{i}$ and 
the known variance-covariance matrix of its measurement errors by $\bm{\Sigma}_{i}$.
Our goal is to use $\bm{X}_{i}$ to estimate  the age, distance modulus, metallicity, and
 zero-age main sequence (ZAMS) mass of the WD.
Our WD model is specified in terms of the $\log_{10}$(age),
distance modulus, metallicity, and ZAMS mass of WDs and
we denote these parameters by ${A}_{i}, {D}_{i}, {Z}_{i}$ and ${M}_{i}$ for $i=1,\cdots, n$
respectively. Because we are primarily interested in WD ages, we group the other stellar parameters
into $\bm{\Theta}_{i}=(D_{i}, Z_{i}, M_{i})$. Finally to simplify notation, we 
write $\bm{X}= (\bm{X}_{i}, \cdots, \bm{X}_{n})$, $A= (A_{1}, \cdots, A_{n})$, and
$\bm{\Theta}= (\bm{\Theta}_{1}, \cdots, \bm{\Theta}_{n})$. Here
we review a case-by-case analysis method for WDs and develop convenient approaches to obtain
the hierarchical modelling fits with improved statistical properties. 

\subsection{Existing Case-by-case Analysis}\label{sec:indv}

The public-domain Bayesian software suite,
Bayesian Analysis of Stellar Evolution with 9 parameters (BASE-9), allows
one to precisely estimate cluster parameters based on photometry
\citep{von2006inverting, degennaro2009inverting, van2009}.
 We have applied BASE-9 to key open clusters \citep{degennaro2009inverting, jeffery2011white, 
hills2015bayesian}, extended BASE-9 to study mass loss from the
main sequence through the white dwarf stage, the so-called Initial Final
Mass Relation (IFMR) \citep{stein2013}, and have demonstrated that BASE-9 can
derive the complex posterior age distributions for individual field white
dwarf stars \citep{o2013}. 

In this article we focus on the development of BASE-9 for fitting the parameters of individual WD stars.
BASE-9 employs a Bayesian approach to fit parameters. 
The statistical model underlying BASE-9 relates a WD's photometry to its parameters,
\begin{equation}\label{eq:singlewd} 
\bm{X}_{i}|A_{i}, \bm{\Theta}_{i}\sim{N}_{l}\Big(G(A_{i}, \bm{\Theta}_{i}), \bm{\Sigma}_{i}\Big),
\end{equation}
where, $N_{l}$ represents a $l$-variate Gaussian distribution,
 $G(\cdot)$ represents the underlying astrophysical models that predicts a star's photometric magnitudes
 as a function of its parameters. Specifically $G$ combines models for the main sequence
through red giant branch \citep[e.g.][]{dotter2008} and the subsequent white dwarf evolution \citep[e.g.][]{bergeron1995, montgomery1999evolutionary}. 

The Bayesian approach employed by BASE-9 requires a 
 joint prior density on $(A_{i}, \bm{\Theta}_{i})$ for each WD. We assume this prior can be factored into
\begin{multline}
{p}(A_{i}, \bm{\Theta}_{i})={p}(A_{i}|\mu_{A_{i}},\sigma_{A_{i}}){p}(D_{i}|\mu_{D_{i}},\sigma_{D_{i}})\times\\
{p}(Z_{i}|\mu_{Z_{i}},\sigma_{Z_{i}}){p}(M_{i}),\label{eq:indvfit}
\end{multline}
where, the individual prior distributions on age, distance modulus, and metallicity
${p}(A_{i}|\mu_{A_{i}},\sigma_{A_{i}})$, 
${p}(D_{i}|\mu_{D_{i}},\sigma_{D_{i}})$,
and ${p}(Z_{i}|\mu_{Z_{i}},\sigma_{Z_{i}})$ are normal densities each with its own prior mean 
(i.e., $\mu_{A_{i}}, \mu_{D_{i}}$, and $\mu_{Z_{i}}$) and standard deviation (i.e., $\sigma_{A_{i}}, \sigma_{D_{i}}$, and $\sigma_{Z_{i}}$). When possible, these prior distributions are specified 
using external studies. The prior on the mass $M_{i}$ is specified as the initial mass function (IMF)
taken from \citet{miller1979}, i.e.,
$\log_{10}(M_{i})\sim{N}(\mu=-1.02, \sigma=0.67729)$. 
BASE-9 deploys a MCMC sampler to separately obtain a MCMC sample from each 
of the WD's joint posterior distributions, 
\begin{multline}\label{eq:jpost}
{p}(A_{i}, \bm{\Theta}_{i}|\bm{X}_{i})
\propto{p}(\bm{X}_{i}|A_{i}, \bm{\Theta}_{i}){p}(A_{i}|\mu_{A_{i}},\sigma_{A_{i}})\times\\
{p}(D_{i}|\mu_{D_{i}},\sigma_{D_{i}}){p}(Z_{i}|\mu_{Z_{i}},\sigma_{Z_{i}})p(M_{i}).
\end{multline}
In this manner, we can obtain case-by-case fits of $A_{i}$ and $\bm{\Theta}_{i}$ for each WD
using BASE-9. 

In this paper for both the case-by-case and the hierarchical analysis, we obtain
 MCMC samples for most of the parameters.  
After we obtain a reasonable MCMC sample for $A_{i}, \bm{\Theta}_{i}, i=1, \cdots, n$,
we estimate these quantities and their $1\sigma$ error bars 
using the means and standard deviations of their MCMC samples,
respectively. For example, letting $A_{i}^{(s)}, s=1, \cdots, S$ be
a MCMC sample for $A_{i}$ of size $S$, after suitable burn-in \citep{degennaro2009inverting},
the posterior mean and standard deviation of $A_{i}$
are approximated by
\begin{align}
&\hat{A}_{i}=\sum_{s=1}^{S}A_{i}^{(s)}/S,\label{eq:postmean} \\
&\hat{\sigma}_{A_{i}}=\sqrt{\sum_{s=1}^{S}(A_{i}^{(s)}-\hat{A}_{i})^{2}/(S-1)}. \label{eq:poststd}
\end{align}
When the posterior distribution of the parameter $A_{i}$ is highly asymmetric, its posterior mean and
$1\sigma$ error bar may not be a good representation of the posterior
distribution.
In this case, we might instead compute
the 68.3\% posterior interval of $A_{i}$ as the range between
the 15.87\% and 84.13\% quantiles of the MCMC sample.

\subsection{Hierarchical Modelling of a Group of WDs}\label{sec:hier}

In this section, we embed the model in Eq. \ref{eq:singlewd} into
a hierarchical model for a sample of
halo WDs,
\begin{equation}\label{eq:hierWD}
\begin{split}
\bm{X}_{i}|A_{i}, \bm{\Theta}_{i}&\sim{N}_{l}\Big(G(A_{i}, \bm{\Theta}_{i}), \bm{\Sigma}_{i}\Big),\\
{A}_{i}&\sim{N}(\gamma,\tau).
\end{split}
\end{equation}
In this hierarchical model $A_{i}, D_{i}, Z_{i}$, and $M_{i}$ are the object-level parameters,
 while $\gamma$ and $\tau$ are population-level parameters, the mean and standard deviation of the $\log_{10}$ ages
  of WDs in the Galactic halo. The assumption of a common population incorporating
  an age constraint is the source of the statistical shrinkage that we illustrate below.
 For the prior distributions of each $\bm{\Theta}_{i}$, we take the same
 strategy as in the case-by-case analysis in Eq. \ref{eq:indvfit}. For the population-level parameters $\gamma$
 and $\tau$, we again choose the uninformative prior distribution, i.e.,
$p(\gamma, \tau)\propto 1$.
 The joint posterior distribution for parameters in the hierarchical model is
 \begin{multline}\label{eq:hjpost}
 {p}(\gamma, \tau, A, \bm{\Theta}|\bm{X})
 \propto{p}(\gamma,\tau)\prod_{i=1}^{n}{p}(\bm{X}_{i}|A_{i}, \bm{\Theta}_{i}){p}(A_{i}|\gamma,\tau) \times\\
 {p}(D_{i}|\mu_{D_{i}},\sigma_{D_{i}}){p}(Z_{i}|\mu_{Z_{i}},\sigma_{Z_{i}}){p}(M_{i}).
 \end{multline}

\subsubsection{Fully Bayesian Method}\label{sec:mhg}

The FB approach obtains a MCMC sample from the joint posterior distribution in Eq. \ref{eq:hjpost}. 
Here we employ a two-stage algorithm \citep{si2017}  to obtain the FB results.
This algorithm takes advantage of
the case-by-case samples in Section \ref{sec:indv} and is easy to implement.
A summary of the computational details of FB appears in Appendix \ref{sec:fb}.

\subsubsection{Empirical Bayes Method}\label{sec:eb}
 We also illustrate how to fit the hierarchical model in Eq. \ref{eq:hierWD} with
 EB. First the joint posterior distribution for $\gamma$ and $\tau$ is calculated as
 \begin{equation}\label{eq:ppost}
 {p}(\gamma, \tau|\bm{X})=\int\cdots\int{p}(\gamma,\tau, A, \bm{\Theta}|\bm{X}){\rm d}{A}{\rm d}\bm{\Theta}.
 \end{equation}
 The integration in Eq. \ref{eq:ppost} is $4\times{n}$ dimensional, which is computationally challenging.
 To tackle this, we use the Monte Carlo Expectation-Maximization (MCEM) algorithm
 \citep[see e.g.,][]{dempster1977, wei1990} to find
 the MAP estimates of $\gamma$ and $\tau$.
 To avoid estimating $\tau$ as zero when its (profile) posterior distribution is highly skewed \citep[e.g.,][]{park2008searching}, 
we again implement EB-log ($\xi=\log\tau$) or EB-inv ($\delta=1/\tau$).
For EB-log, the joint posterior distribution of $\gamma$ and $\xi$ equals
 $${p}(\gamma,\exp(\xi)|\bm{X})\exp(\xi),$$
where $p(\cdot| \bm{X})$ is the joint posterior distribution of $\gamma$ and $\tau$.
The EB-log method proceeds in two steps.
\begin{description}
 \item[\textbf{Step 1:}] Deploy MCEM to obtain the MAP estimates of $\gamma$ and $\xi$, and 
 transform to $\gamma$ and $\tau$, i.e.,
 \begin{equation}\label{eq:eb.mcem1}
(\hat{\gamma},\hat{\tau})=\arg\max_{\gamma, \tau}{p}(\gamma, \tau|\bm{X})\tau.
\end{equation}
For details of MCEM in this setting, see Appendix \ref{sec:mcem}. 
 \item[\textbf{Step 2:}] For WD $i=1, \cdots, n$, we obtain a MCMC sample from 
 $${p}(A_{i}, \bm{\Theta}_{i}|\bm{X}_{i}, \hat{\gamma}, \hat{\tau})\propto{p}(\bm{X}_{i}|A_{i}, \bm{\Theta}_{i})
 {p}(\bm{\Theta}_{i}){p}(A_{i}|\hat{\gamma}, \hat{\tau})$$
 using BASE-9.
 \end{description}
 EB-inv proceeds in a similar manner, but with Eq. \ref{eq:eb.mcem1} replaced with
 $$(\hat{\gamma},\hat{\tau})=\arg\max_{\gamma, \tau}{p}(\gamma, \tau|\bm{X})\tau^{2},$$
where $p(\cdot| \bm{X})$ is again the posterior distribution of $\gamma$ and $\tau$.

\section{Simulation Study}\label{sec:simulation}

To illustrate the performance of the various estimators of the object-level 
WD ages and the population-level parameters $\gamma$ 
and $\tau$, we perform a set of simulation studies.
Because the relative advantage of the shrinkage estimates compared with the case-by-case estimates
depends both on the precision of the case-by-case estimates and the degree of
heterogeneity of the object-level parameters, we repeat the simulation study under five
scenarios, each with different values of observation error matrix $\bm{\Sigma}$  and population standard deviation 
of $\log_{10}$(age), i.e., of $\tau$. 
We simulate the parameters $\{A_{i}, D_{i}, T_{i}, M_{i}, i=1,2, \cdots, N_{1}\}$ for each group of WDs 
from the distributions in Table \ref{tab:dist}, where $\gamma=10.09$ (12.30 Gyr) is the population mean and $\tau$
varies among the simulation settings given in Table \ref{table:summary1}. 
For consistency with the data analyses in  Section \ref{sec:real},
we simulate $u, g, r, i, z, J, H, K$ magnitudes for all WDs. 
Using BASE-9 for each setting,
we simulate $N_{2}=25$ replicate datasets, each composed of $N_{1}=10$ halo WDs. 
For each WD in every group, we generate its $\log_{10}$(age), distance modulus,
metallicity and mass from distributions in Table \ref{tab:dist}, where $\tau$
is given in Table \ref{table:summary1}.
The particular values and truncations in Table \ref{tab:dist} and \ref{table:summary1} are chosen because they reflect plausible values for
 actual halo WDs.
 \begin{table*}
 \centering
 \caption{Distributions used to simulate the parameter values.\label{tab:dist}}
 \begin{tabular}{l l}
 \hline\hline
 Parameters &  Distributions \\ \hline
 $\log_{10}$(age) &  ${N}(\gamma = 10.09, \tau)$ truncated between $9.7$ to $10.17$ \\
 Distance modulus &  ${N}(4.0 , 2.5^{2})$ truncated to interval $(2.7, 5.7)$, i.e. 34 to 138 pc   \\
 Metallicity & ${N}(-1.5, 0.02)$ \\ 
 Mass  & $\log_{10}(\text{Mass})\sim{N}(-1.02, 0.67729)$ truncated to $(0.8, 3.0)$  \\
 \hline
 \end{tabular}
  \end{table*}
 
\begin{table*}
\begin{center}
\caption{Comparing the statistical properties of the various shrinkage and case-by-case
estimates under five simulated settings.\label{table:summary1}}
\begin{tabular}{ll llc llc llc llc ll}
\hline\hline
\multicolumn{2}{l}{\multirow{2}{*}{Simulation Settings}} & \multicolumn{3}{c}{EB} & \multicolumn{3}{c}{EB-log} & \multicolumn{3}{c}{EB-inv} & \multicolumn{3}{c}{FB} & \multicolumn{2}{c}{Case-by-case} \\  \cline{3-4}\cline{6-7}\cline{9-10}\cline{12-13}\cline{15-16}
\multicolumn{2}{l}{}                                  & MAE       & RMSE       && MAE         & RMSE         && MAE         & RMSE         && MAE       & RMSE       && MAE            & RMSE           \\\hline
 \multicolumn{2}{l}{I:~~~$\tau=0.05, \bm{\Sigma}=\bm{\Sigma}_{0}$}  & 0.028 & 0.038 && 0.029 & 0.042 && 0.033 & 0.052 && 0.032 & 0.047 && 0.084 & 0.18\\ \hline
\multicolumn{2}{l}{II:~~$\tau=0.03, \bm{\Sigma}=0.8^{2}\bm{\Sigma}_{0}$}    &0.024 & 0.035 && 0.025 & 0.038 && 0.027 & 0.044 && 0.025 & 0.041 && 0.086 & 0.19     \\
   \multicolumn{2}{l}{III:~$\tau=0.03, \bm{\Sigma}=1.2^{2}\bm{\Sigma}_{0}$}    & 0.023 & 0.031 && 0.024 & 0.034 && 0.026 & 0.037 && 0.025 & 0.035 && 0.099 & 0.20        \\ \hline
 \multicolumn{2}{l}{IV:~~$\tau=0.06, \bm{\Sigma}=0.8^{2}\bm{\Sigma}_{0}$}    & 0.038 & 0.063 && 0.038 & 0.070 && 0.040 & 0.074 && 0.034 & 0.049 && 0.12 & 0.26  \\ 
        \multicolumn{2}{l}{V:~~~$\tau=0.06, \bm{\Sigma}=1.2^{2}\bm{\Sigma}_{0}$}    & 0.033 & 0.046 && 0.032 & 0.044 && 0.032 & 0.045 && 0.032 & 0.044 && 0.10 & 0.22\\ \hline             
\end{tabular}
\end{center}
\end{table*}

We compute the empirical
 standard error for each simulated magnitude by averaging the errors from the observed halo WDs in Section \ref{sec:real},  
  and we denote by $\bm{\Sigma}_{0}$ the variance matrix of observed magnitudes, i.e., the square of
  empirical standard errors for all eight magnitudes. Specifically, $\bm{\Sigma}_{0}$ is a 
  diagonal matrix with diagonal elements equal to $(0.304^{2}, 0.092^{2}, 0.027^{2}, 0.026^{2}, 0.068^{2},
  0.062^{2}, 0.086^{2}, 0.083^{2})$.
  
For simplicity in each setting 
all stars share the same diagonal observation variance, that is each $\bm{\Sigma}_{i}=\bm{\Sigma}, i =1, 2, \cdots, N_{1}$. The observation error variances for five simulation settings are described in terms of $\bm{\Sigma}_{0}$ in Table \ref{table:summary1}. 
In the entire simulation study, we
employ the \citet{dotter2008} WD precursor models, 
\citet{renedo2010new} WD interior models, \citet{bergeron1995} WD atmospheres
and \citet{williams2009} IFMR.

Subsequently we recover parameters with multiple approaches:
EB, EB-log, EB-inv, FB, and the case-by-case analysis.
We specify
non-informative broad prior distributions on each $\bm{\Theta}_{i}$, namely
$D_{i}\sim{N}(4.0, 2.4^{2})$, $Z_{i}\sim{N}(-1.5, 0.5^{2})$ and $\log_{10}(M_{i})\sim{N}(-1.02,0.67729)$.
The case-by-case analyses requires a
prior distribution on each $A_{i}$ and we use ${p}(A_{i})\propto 1$. 
The hierarchical model in Eq. \ref{eq:hierWD} requires priors on $\gamma$
and $\tau$, and we again use ${p}(\gamma, \tau)\propto 1$.
We compare the case-by-case estimates with shrinkage estimates based on
the hierarchical model.
Results from the case-by-case analyses (obtained by fitting the model in Eq. \ref{eq:singlewd})
are indicated 
with a superscript $I$ (for ``individual'') and those from hierarchical analyses 
(obtained by fitting to model in Eq. \ref{eq:hierWD}) 
are indicated with an $H$.

We denote $\log_{10}$(age) of the $i$-th simulated WD in the $j$-th replicate group by $A_{ij}$. 
Using both the case-by-case and hierarchical analyses, we obtain MCMC samples of
the parameters $A_{ij}, i=1, 2, \cdots, N_{1}, j=1, 2, \cdots, N_{2}$.
We estimate $A_{ij}$ by taking the MCMC sample mean as in Eq. \ref{eq:postmean}
and denote the estimates based on case-by-case and hierarchical
analyses by $\hat{A}_{ij}^{I}$ and $\hat{A}_{ij}^{H}$, respectively. We
compute the absolute value of the error\footnote{We use this term to refer to the absolute value of
the error.} of each estimator $\hat{A}_{ij}$ as
$${\rm error}(\hat{A}_{ij})=|\hat{A}_{ij}-{A}_{ij}|.$$
In our simulation study, we are mainly concerned with the difference between absolute
 errors from shrinkage and case-by-case estimates 
 \begin{align*}
{\rm Diff}(A_{ij}) &={\rm error}(\hat{A}_{ij}^{I})-{\rm error}(\hat{A}_{ij}^{H})\\
&=|\hat{A}^{I}_{ij}-{A}_{ij}|-|\hat{A}^{H}_{ij}-{A}_{ij}|,
\end{align*}
which compares the prediction accuracy of the two methods.
If ${\rm Diff}(A_{ij})\geq0$, then 
 the absolute deviation of the case-by-case estimate of star $A_{ij}$ is greater than that of the shrinkage estimate. 
 
\begin{figure*}
\centering
\includegraphics[width=0.7\textwidth]{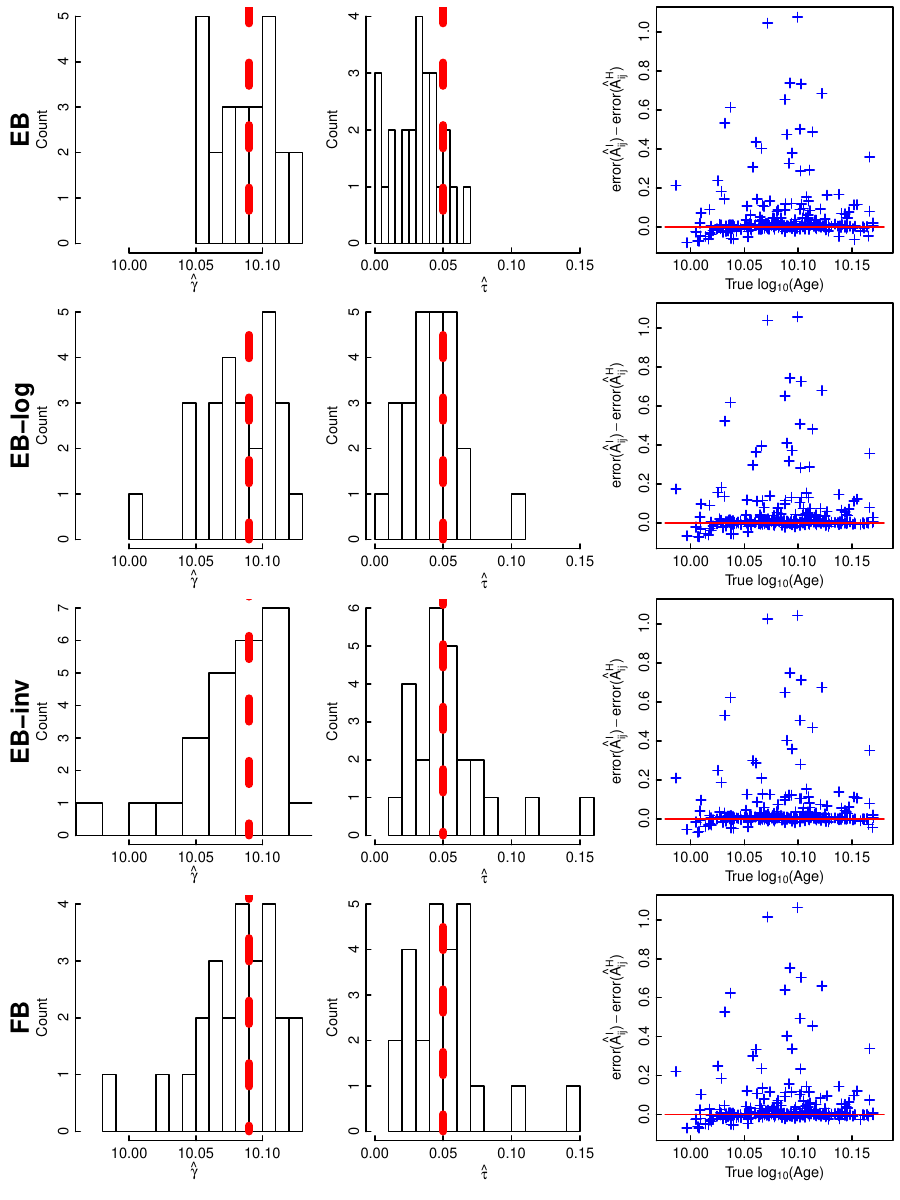}
\caption{Results for Simulated Setting I, with $\tau=0.05$ (standard deviation of $\log_{10}$(age)) and $\Sigma=\Sigma_{0}$ (photometric error variances).
The rows correspond to various fitting methods:
EB, EB-log, EB-inv and FB, respectively. Each row includes: i) a histogram of the
fitted values of $\hat{\gamma}$ across the $N_{2}=25$ simulation replicates, ii) a
histogram of $\hat{\tau}$, and iii) a comparison of the difference in absolute errors in
the fitted WD ages obtained with the
case-by-case analysis versus each of the hierarchical methods.
The thick dashed red lines in each histogram indicate the true values of $\gamma$ or $\tau$.
  In the right-most column, positive vertical values indicate larger error for the
 case-by-case fit and the thin horizontal lines correspond to equal errors for
 the two methods. Figures \ref{hist2}--\ref{hist5} for Simulated Settings II--V are shown in the online supplementary material.}
 \label{hist1}
\end{figure*}
Fig.~\ref{hist1} compares the performance of shrinkage estimates under simulated Setting I ($\tau=0.05, \bm{\Sigma}=\bm{\Sigma}_{0}$). The corresponding summary plots for the other simulation settings are
similar and appear in Fig.s \ref{hist2}--\ref{hist5} the online supplement.
The histograms in Fig. \ref{hist1} demonstrate that
the estimates of $\gamma$ and $\tau$ are generally close to their true values (thick, dashed red lines). 
Under all $5$ settings, however, for some replicate groups of halo WDs,
EB produces estimates of $\tau$ equal to $0$, see the first row, middle panel of Fig. \ref{hist1}. (This
phenomenon is fully discussed in Appendix \ref{sec:lmc} and Fig. \ref{fig:profile}.)
In these cases, the shrinkage estimate of the age of each WD in these are equal,
which potentially leads to large errors.
As we mentioned in Section \ref{sec:eb}, this highlights a difficulty with EB, and
demonstrates the need for the transformed EB-log or EB-inv.
Both of
these approaches produce similar results to EB, but avoid the possibility of $\hat{\tau}=0$.
The third column in Fig. \ref{hist1} shows the scatter plot of
${\rm Diff}(A_{ij})$ against $A_{ij}$, $i=1, 2, \cdots, N_{1}, j=1, 2, \cdots, N_{2}$.
Because most of the scatter in these plots is above the solid red zero line,
the estimates of $\log_{10}$(age) from the case-by-case 
analyses tend to be further from the true values
than the shrinkage estimates. Approximately two thirds of the $N_{1}\times{N}_{2}$ simulated stars in each setting are
better estimated with the shrinkage method than the case-by-case fit.
For stars below the red solid lines, nominally the case-by-case fit is better, but the advantage is small. In fact,
for almost all simulated stars,
${\rm Diff}(A_{ij})={\rm error}(A_{ij}^{I})-{\rm error}(A_{ij}^{H}) > -0.1$,
so the shrinkage estimates do not perform much worse than case-by-case estimates for
any WD and often perform much better.
For some stars, we have ${\rm Diff}(A_{ij}) >0.5$.
From the point of view of reliability of the technique, it is comforting that 
the four hierarchical fits (EB, EB-log, EB-inv, FB) perform similarly, at least
when $\hat{\tau}>0$ for EB.

\begin{table}
\begin{center}
\caption{The percentage of WDs with improved age estimates.\label{table:percent}}
\begin{tabular}{ll llll}
\hline\hline
\multicolumn{2}{l}{Simulation Settings} &EB & EB-log & EB-inv & FB\\ \hline
  $\tau=0.05$ & $\bm{\Sigma}=\bm{\Sigma}_{0}$  & 65.6\% & 69.6\% & 68.8\% & 66.8\%\\ \hline
\multirow{2}{*}{$\tau=0.03$}    & $\bm{\Sigma}=0.8^{2}\bm{\Sigma}_{0}$    & 74.4\% & 72.8\% & 76.4\% & 73.2\% \\
                       & $\bm{\Sigma}=1.2^{2}\bm{\Sigma}_{0}$    & 67.6\% & 69.6\% & 70.4\% & 72.8\%      \\ \hline
\multirow{2}{*}{$\tau=0.06$}    & $\bm{\Sigma}=0.8^{2}\bm{\Sigma}_{0}$    & 65.6\% & 67.6\% & 68.0\% & 64.4\% \\ 
                          & $\bm{\Sigma}=1.2^{2}\bm{\Sigma}_{0}$    & 60.4\% & 64.0\% & 66.8\% & 63.2\%\\ \hline             
\end{tabular}
\end{center}
\end{table}

Table \ref{table:summary1} presents a
 numerical comparisons of the shrinkage and case-by-case estimates of $\log_{10}$(age).
Specifically it presents the average mean absolute error (MAE) and the average root of the mean squared error (RMSE)
for each method, i.e.,
\begin{align*}
&{\rm MAE}({A})=\frac{1}{N_{1}\cdot{N}_{2}}\sum_{j=1}^{N_{2}}\sum_{i=1}^{N_{1}}|\hat{A}_{ij}-{A}_{ij}|,\\
&{\rm RMSE}({A})=\sqrt{\frac{1}{N_{1}\cdot{N}_{2}}\sum_{j=1}^{N_{2}}\sum_{i=1}^{N_{1}}(\hat{A}_{ij}-{A}_{ij})^{2}}.
\end{align*}
Both MAE and RMSE measure the distance between the true values and their estimates. 
Smaller MAE and RMSE indicates that the estimate is more accurate.
Table \ref{table:summary1} summarises the performance of different estimators
under the five simulated settings.
In terms of MAE and RMSE, all four shrinkage estimates (EB, EB-log, EB-inv, FB) 
are significantly better than the case-by-case estimates,
though there are slight differences among the four shrinkage estimates.
Table \ref{table:percent} reports the percentage of simulated WDs that are better estimated by shrinkage methods
than the case-by-case fits for each of the four statistical approaches and each of the five
simulation settings.
 We conclude that 60\%--75\% of simulated stars have a more reliable age estimate 
 from the hierarchical analyses than from the case-by-case analyses.

From Tables \ref{table:summary1} and \ref{table:percent}, we conclude that shrinkage estimates from both EB-type
and FB approaches outperform the case-by-case analyses in terms of smaller RMSE and MAE. Under the five
simulated settings, all four computational methods, EB, EB-log, EB-inv and FB, behave similarly. Their MAEs
and RMSEs are comparable. Also, the percentages of better estimated WDs from these four approaches are  
consistent. 

Simulation Setting III ($\tau=0.03, \bm{\Sigma}=1.2^{2}\bm{\Sigma}_{0}$) benefits most from the
shrinkage estimates in terms 
of reduced RMSE. The RMSE from the case-by-case fits under Simulation Setting III is approximately 0.20, while
the RMSE from the EB is around $0.031$, less than one sixth of the former. 
Simulation Setting IV ($\tau=0.06, \bm{\Sigma}=0.8^{2}\bm{\Sigma}_{0}$) gains the least from the shrinkage estimates; the RMSE of EB ($0.063$) is
about a quarter of the RMSE of case-by-case ($0.26$). Generally, when $\Sigma$ is large and $\tau$ is small,
the advantage of shrinkage estimates is the greatest.
With small $\Sigma$ and large
$\tau$, the advantage of shrinkage estimates over case-by-case estimates decreases. This is consistent to statistical
theory \citep[see][Chapter 5]{gelman2014}.
Generally speaking, using EB-log rather than EB to avoid a fitted variance of zero. In terms
of computational investment, the FB algorithm is less time-consuming than all of our EB algorithms.

\section{Analysis of a Group of Candidate Halo White Dwarfs}\label{sec:real}

 Now we turn to the hierarchical and case-by-case
analysis of the $10$ field WDs from the Galactic halo listed in 
Table~\ref{table:prior}.
 In the entire analysis, we
employ the \citet{dotter2008} WD precursor models, 
\citet{renedo2010new} WD interior models, \citet{bergeron1995} WD atmospheres,
and \citet{williams2009} IFMR.
 
We acquire prior densities on ${M}_{i}, {D}_{i}$, and $Z_{i}, i=1, \cdots, 10$ from 
the literature \citep{kilic2010, kilic2012, gianninas2015ultracool, dame2016new}. The
atmospheric compositions and priors on distance moduli for
these $10$ stars  are listed in Table~\ref{table:prior}. We use
a ZAMS mass prior IMF from \citet{miller1979} on $M_{i}$ and
a diffuse  prior on metallicity $Z_{i}\sim{\rm N}(-1.50, 0.5)$.
In this article, we do not leave the WD core composition as a free parameter,
but instead, we use the WD cooling model derived from the work of \citet{renedo2010new}.

 \begin{table}
\begin{center}
\caption{Prior distributions for distance moduli and atmosphere composition for halo WD sample\label{table:prior}}
\begin{tabular}{l c c l}
\hline\hline
White Dwarf       & Dist. Mod.  & Comp. & Reference  \\
                        &         $(V-M_{V})$                        &          &\\\hline
J0301$-$0044 & ${\rm N}(4.35, 0.2)^{a}$ &He  & Paper \textrm{III}$^{d}$ \\
J0346$+$246   & ${\rm N}(2.24, 0.33)^{a}$    &   H    & Paper \textrm{II}$^{c}$ \\ 
J0822$+$3903 & ${\rm N}(5.19, 2.4)$   & H& Paper \textrm{IV}$^{e}$\\
J1024$+$4920 & ${\rm N}(5.54, 2.4)$    &He& Paper \textrm{IV}$^{e}$\\
J1102$+$4113    & ${\rm N}(2.64, 0.13)^{a}$  &H& Paper \textrm{II}$^{c}$ \\ 
J1107$+$4855 & ${\rm N}(3.41, 0.19)^{a}$     &H& Paper \textrm{III}$^{d}$ \\
J1205$+$5502 & ${\rm N}(5.04, 2.4)$     &He& Paper \textrm{IV}$^{e}$\\
J2137$+$1050  & ${\rm N}(4.46, 2.4)$           &H& Paper \textrm{I}$^{b}$\\ 
J2145$+$1106N  & ${\rm N}(4.19, 2.4)$           &H& Paper \textrm{I}$^{b}$\\ 
J2145$+$1106S   & ${\rm N}(4.23, 2.4)$           &H& Paper \textrm{I}$^{b}$\\ \hline
\multicolumn{4}{l}{$^a$ Precise distance moduli are from trigonometric parallax}\\
\multicolumn{4}{l}{measurements.}\\
\multicolumn{4}{l}{$^b$ Paper \textrm{I} is \citet{kilic2010}.}\\
\multicolumn{4}{l}{$^c$ Paper \textrm{II} is \citet{kilic2012}.}\\
\multicolumn{4}{l}{$^d$ Paper \textrm{III} is \citet{gianninas2015ultracool}.}\\
\multicolumn{4}{l}{$^e$ Paper \textrm{IV} is \citet{dame2016new}.}\\
 \end{tabular}
\end{center}
\end{table}

 For the case-by-case fitting of each WD,
we employ a minimally informative flat prior on $A_{i}$, specifically, 
 $A_{i}\sim{\rm Unif}(8.4, 10.17609)$.

\begin{figure*}
\centering
\includegraphics[width=\textwidth]{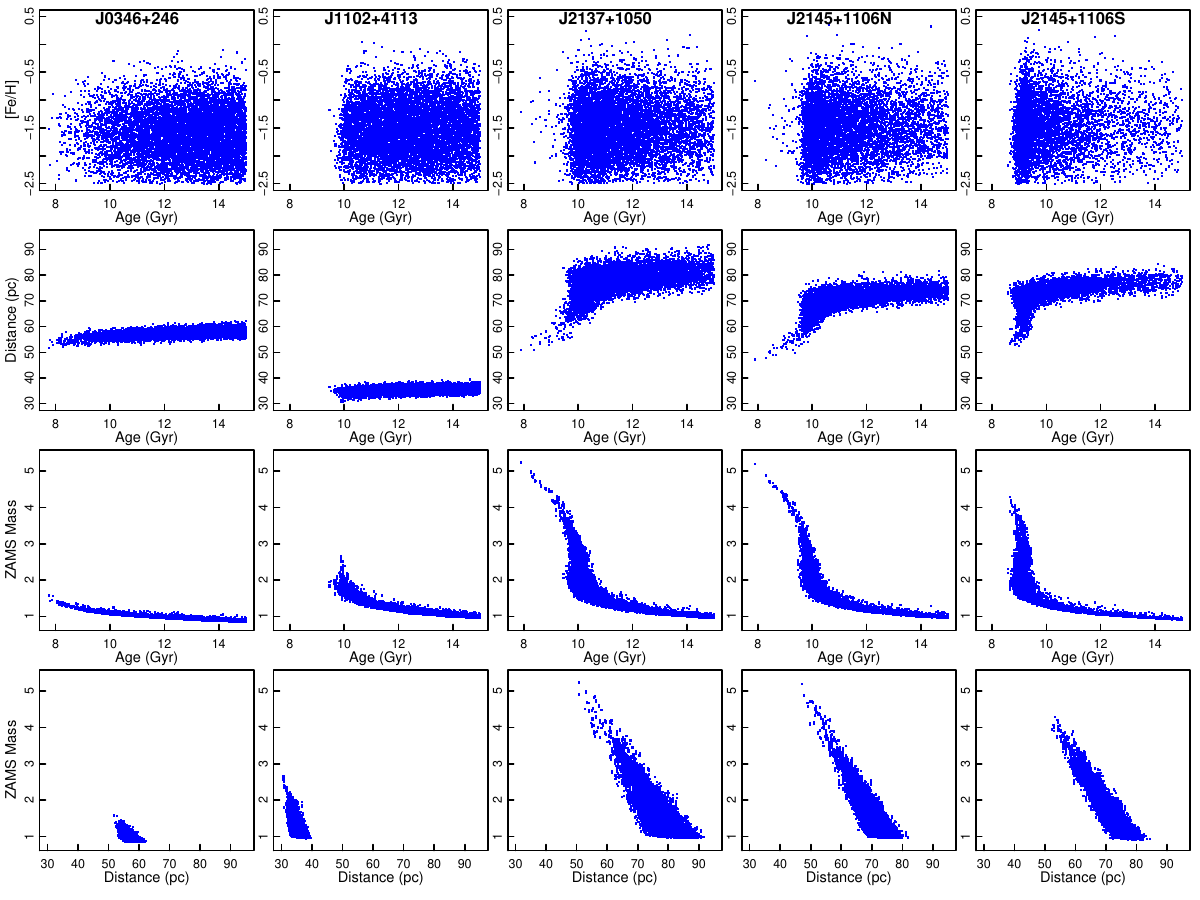}
\caption{Case-by-case results: projections of the joint posterior distributions onto the two dimensional 
planes of (from top to bottom) age--metallicity, age--distance, age--ZAMS mass, and distance--ZAMS mass
 for five of the Galactic halo WDs (columns). \label{fig:wds}}
\end{figure*}

\subsection{Case-by-case Analysis}
We derive the joint posterior density for the parameters using Bayes' theorem:
\begin{align}
{p}(A_{i}, D_{i}&, Z_{i}, M_{i}|\bm{X}_{i})\propto\nonumber\\
&{p}(\bm{X}_{i}|A_{i}, D_{i}, Z_{i}, M_{i}){p}(A_{i}){p}(D_{i}){p}(Z_{i}){p}(M_{i}).\label{singlepost}
\end{align}
Before specifying a hierarchical modelling for the $10$ WDs,
 we obtain their case-by-case fits using BASE-9 \citep[as in][]{o2013}. 
By using the priors in Table ~\ref{table:prior} and as described above, we fit each of $10$ halo WDs 
individually with 
BASE-9. We present results for $5$ typical stars in Fig. \ref{fig:wds}.

Each column in Fig.~\ref{fig:wds} corresponds to one WD. The rows
provide different two dimensional projections of the posterior distributions. The
asymmetric errors in the fitted parameters, including age, are evident.
 The first
row illustrates that the correlation  between the metallicity and age for these five WDs is weak.
In the second row, the distance and age of WDs have a strong positive correlation for ages less
than 10 Gyrs. However, this pattern generally disappears 
for ages greater than 10 Gyrs. From the third and fourth 
row, the ZAMS mass displays a clear negative correlations with both age and distance.

The plot shows that
the range of possible ages for these five stars is large, from $8$ Gyrs to $15$ Gyrs.
Assuming each of these ten WDs are bona fide Galactic halo members, we expect their ages to be similar.
However, their masses, distance moduli and metallicities may vary substantially.
  In this situation, it is sensible to deploy hierarchical modelling on the ages of these $10$ WDs, 
  which provides substantial additional information.
\begin{figure*}
\centering
\includegraphics[width=\textwidth]{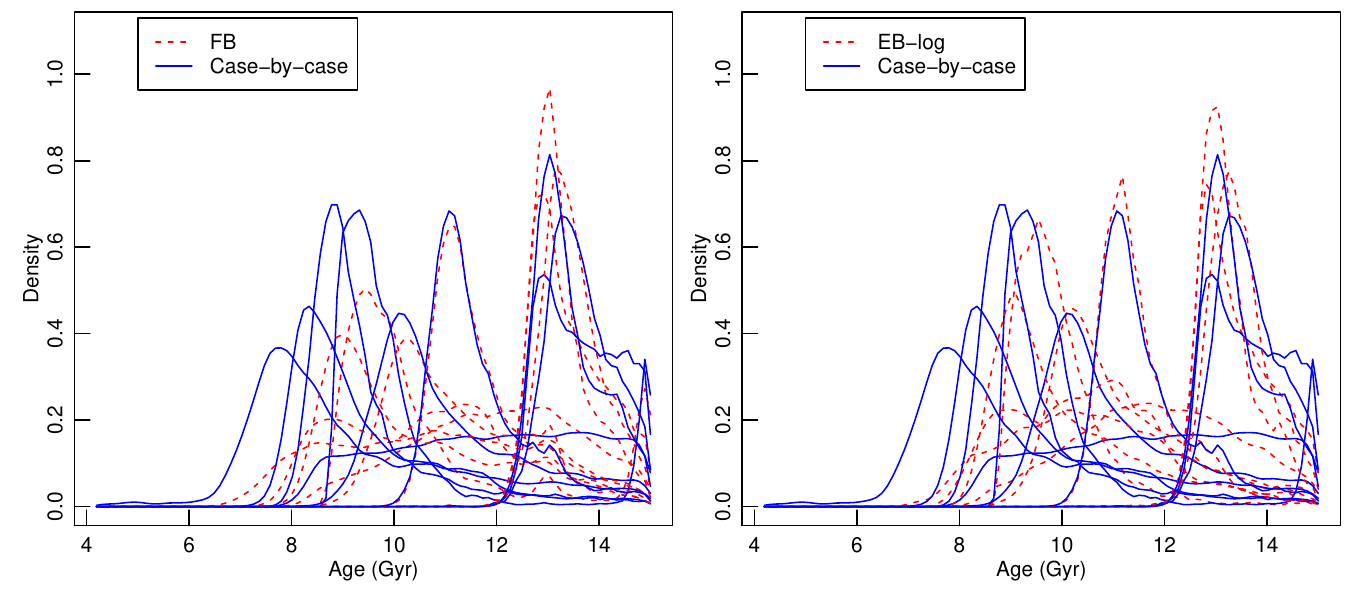}
\caption{Comparison of the posterior distributions of ages of WDs obtained with the
hierarchical analyses and with the case-by-case analysis.
The left panel uses the FB approach to fit the hierarchical model, whereas the right panel uses EB-log.
Both hierarchical approaches give age estimates that are more consistent with each other than
with the case-by-case estimates.}
\label{fig:hierComp}
\end{figure*}
\begin{table*}
\centering
\caption{Estimates of the ages (in Gyr) of ten candidate halo WDs.}
\begin{tabular}{lllllllll}
\hline
\multirow{2}{*}{WD name} & \multicolumn{2}{c}{FB}                  &  & \multicolumn{2}{c}{EB-log}     &  & \multicolumn{2}{c}{Case-by-case} \\ \cline{2-3} \cline{5-6} \cline{8-9} 
                     & Mean   & Posterior Interval &  & Mean   & Posterior Interval &  & Mean      & Posterior Interval    \\ \hline
J0301$-$0044           & 11.97  & (10.26, 13.65) && 11.85 & (10.24, 13.5) & &11.84 & (9.6, 14)      \\
J0346$+$246            & 10.68 & (9.24, 12.97) & &9.92 & (9.18, 10.56) && 10.23 & (9.07, 10.88)     \\
J0822$+$3903           & 11.33 & (10, 12.91) & &11.12 & (9.97, 12.43)   && 11.1 & (9.81, 12.74) \\ 
J1024$+$4920           & 10.65 & (8.57, 12.5) && 10.55 & (8.98, 11.9) && 8.94 & (7.43, 10.9) \\
J1102$+$4113           & 13.41 & (12.74, 14.23) && 13.41 & (12.74, 14.22) && 13.64 & (12.84, 14.53)    \\
J1107$+$4855           & 10.4 & (8.82, 12.48) && 10.07 & (8.83, 11.51) && 9.46 & (8.53, 10.35)     \\
J1205$+$5502           & 10.89 & (8.78, 13.01) & &10.67 & (8.88, 12.45) && 9.77 & (8.22, 11.83)    \\
J2137$+$1050           & 13.47 & (12.93, 14.08) && 13.46 & (12.91, 14.06) && 13.63 & (13.03, 14.35)     \\
J2145$+$1106N          & 13.25 & (12.74, 13.84) && 13.24 & (12.73, 13.81) && 13.4 & (12.8, 14.15)     \\
J2145$+$1106S          & 11.68 & (10.87, 12.71) && 11.54 & (10.83, 12.37) && 11.65 & (10.84, 12.64)     \\ \hline
\end{tabular}
\label{tab:wds}
\end{table*}
\begin{figure*}
\centering
\includegraphics[width=0.7\textwidth]{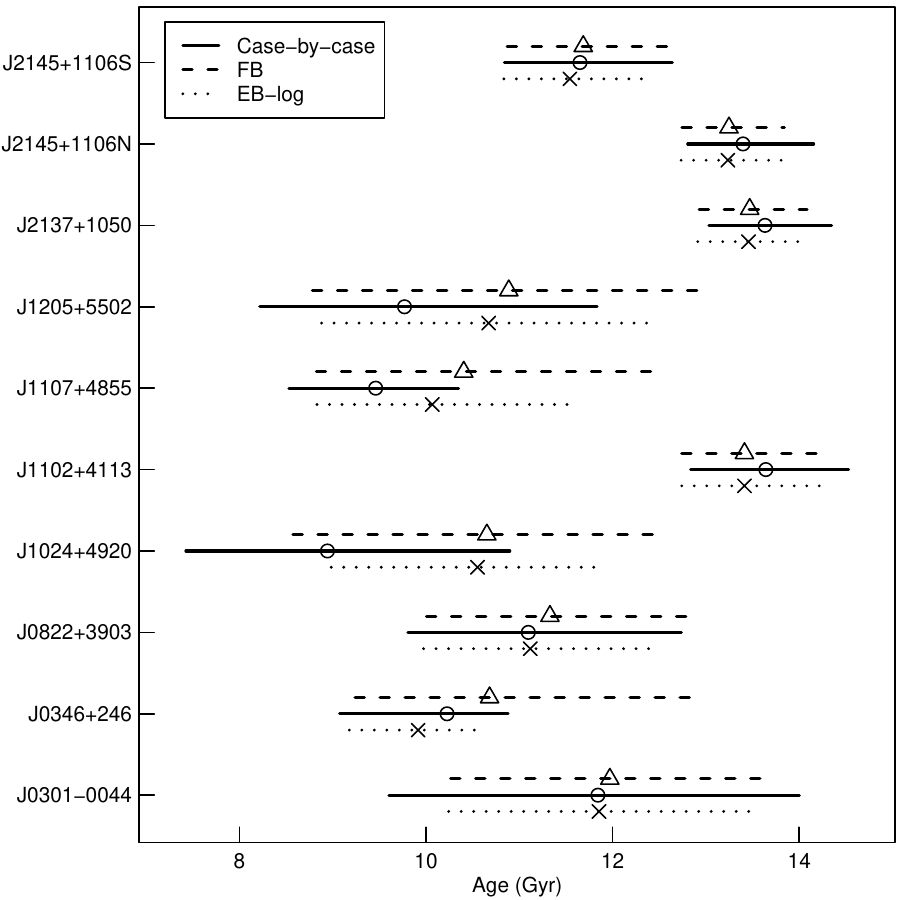}
\caption{The EB-log, FB, and case-by-case estimates of the ages of the ten candidate halo WDs. The interval estimates
are central $68.3\%$ posterior intervals of the posterior distributions. The dashed, solid, and dotted intervals are
computed from FB, case-by-case, and EB-log, respectively. The triangle, circle and cross signs are
the posterior means of the ages for each approach. The hierarchical age estimates of J1205+5502 and J1024+4920
in particular are shrunk towards the age estimates of the other WDs.
\label{fig:ages}}
\end{figure*}

\subsection{Hierarchical Analysis}
Here we deploy both EB-log and FB to
obtain fits of the hierarchical model in Eq. \ref{eq:hierWD} based on ten
candidate Galactic halo WDs. 
In Fig.~\ref{fig:hierComp}, we compare the posterior density distributions of the age of each WD obtained
with the case-by-case method and with that obtained with both EB-log and FB.

Fig.~\ref{fig:hierComp} demonstrates that the posterior distributions of the ages under the
 hierarchical model -- both EB-log and FB -- peak near a sensible halo age,
whereas the case-by-case estimates (solid lines) disperse over a much wider range.
Both EB-log and FB estimates
are consistent, which we discuss further below.
\begin{table*}
\centering
\caption{The 68.3\% posterior/predictive intervals for the age of Galactic halo WDs}
\label{tab:ages}
\begin{tabular}{cllll}
\hline
Items                      & Units     & FB              & EB-log             \\\hline
\multirow{2}{*}{\begin{tabular}[l]{@{}l@{}}Posterior interval for\\the mean age of halo WDs\end{tabular}} & $\log_{10}$(Year)  & $10.07_{-0.04}^{+0.03}$   &            $\text{n.a.}$        \\
          & Gyr &  $12.11_{-0.86}^{+0.85}$              &      $\text{n.a.}$             \\\hline
\multirow{2}{*}{\begin{tabular}[l]{@{}l@{}}Posterior interval for the standard\\deviation of ages of halo WDs\end{tabular}} & $\log_{10}$(Year)  & $0.05\pm0.03$   &            $\text{n.a.}$        \\
          & Gyr &  $1.18^{+0.57}_{-0.62}$              &      $\text{n.a.}$             \\\hline
\multirow{2}{*}{\begin{tabular}[l]{@{}l@{}}Predictive interval for the\\age distribution of halo WDs$^{a}$\end{tabular}}     & $\log_{10}$(Year)  & $10.08\pm0.05$   & $10.06\pm0.07$      \\
                                     & Gyr   &  $12.11_{-1.53}^{+1.40}$            & $11.43\pm1.86$ \\\hline
\multicolumn{4}{l}{$^a$ The 68.3\% predictive interval for the ages of WDs in the Galactic is an estimate of}\\
 \multicolumn{4}{l}{an interval that contains 68.3\% of halo WD ages, taking account of uncertainties}\\
\multicolumn{4}{l}{of both population-level parameters, $\gamma$ and $\tau$, and of the variability}\\
\multicolumn{4}{l}{in the ages of halo WDs.}\\
\end{tabular}
\end{table*}

The photometric errors of these ten WDs are close to $\Sigma_{0}$ in the simulation study. 
So the data is similar to Simulation Setting I ($\tau=0.07, \bm{\Sigma}=\bm{\Sigma}_{0}$).
Hence, the advantage of the shrinkage estimates over the case-by-case estimates shown
 in Simulation Setting I
should be predictive and we expect that the 
hierarchical fits (dotted lines) in Fig. \ref{fig:hierComp} are better estimates of the true ages of
these halo WDs. 

Table \ref{tab:wds} and Fig. \ref{fig:ages} summarise the estimated ages. 
The 68.3\% posterior intervals of 
ages of WDs from EB-log and FB are generally narrower than those from the case-by-case analyses, which
means that shrinkage estimates (FB and EB-log) produce more precise estimates.
The fits and errors from EB-log and FB are quite consistent.

\begin{figure*}
\centering
\includegraphics[width=0.7\textwidth]{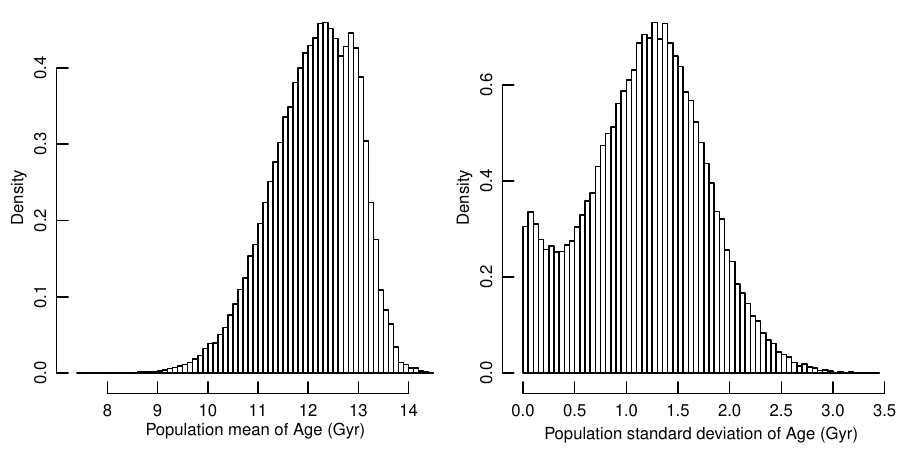}
\caption{Histograms of MCMC samples from the posterior distribution of the
 mean and standard deviation of the population of halo WD ages on the Gyr scale,
 obtained using FB fit to ten candidate
Galactic halo WDs.\label{fig:fbpar}}
\end{figure*}

In both BASE-9 and the hierarchical model (Eq. \ref{eq:hierWD}), the ages, $A_{i}$, of stars are 
specified on the $\log_{10}$(Year) scale. Given an MCMC sample from the posterior
distribution of age on the $\log_{10}$(Year) scale, we can obtain a MCMC sample on the 
age scale by backwards transforming each value in the sample via
\begin{equation}\label{eq:exp1}
\text{age}=10^{(A_{i}-9)},
\end{equation}
where the units for age and $\log_{10}$(age) are Gyr and $\log_{10}$(Year), respectively.
For the population-level parameters $\gamma$ and $\tau$, however,
 the transformation from the $\log_{10}$(Year) scale to the Gyr scale is more complicated. Again, starting
 with the MCMC sample of $\gamma$ and $\tau$, for each sampled pair,
  we (i) generate a Monte Carlo sample of $A_{i}$, (ii) transform this sample
 to the Gyr scale as in Eq. \ref{eq:exp1}, and (iii) compute the mean and standard deviation
of the transformed Monte Carlo sample.
Histograms of the resulting sample from the posterior distribution of the mean
and standard deviation of the age on the Gyr scale appear
 in Fig. \ref{fig:fbpar}. 

We present estimates of the population distribution of the age of Galactic halo WDs in Table \ref{tab:ages},
 on both the Gyr and $\log_{10}$(Year) scales.
 In the first two rows, we report the 68.3\% posterior intervals for the mean ($\gamma$)
and standard deviation ($\tau$) of the distribution of ages of halo WDs, $12.11_{-0.86}^{+0.85}$ Gyr and
$1.18_{-0.62}^{+0.57}$ Gyr, respectively.
The point estimates of the population mean (11.43 Gyr) and standard deviation (1.86 Gyr) 
from EB-log are quite consistent to results of FB.  
EB-log does not directly provide error estimates for the population mean and standard deviation,
but bootstrap techniques \citep{efron1979} could be used. 
We do not pursue this here, because it is computationally expensive and uncertainties
are provided by FB.

In Table \ref{tab:ages}, we also report the 68.3\% predictive intervals of
the age distribution from FB and EB-log,
 which summarises the underlying distribution of halo WDs ages.
 These are our estimates of an interval that contains the ages of
 68.3\% of halo WDs.
From FB, the 68.3\% predictive interval for the distribution of 
halo WD ages is $12.11_{-1.53}^{+1.40}$ Gyr.
In other words, we predict that 68.3\% of WDs in the Galactic halo have ages between
 $10.58$ and $13.51$ Gyr. The 68.3\% predictive interval from EB-log is 
 $11.43\pm1.86$ Gyr, slightly broader than that from the FB.

In summary, our hierarchical method finds that the Galactic halo has a mean age of
$12.11_{-0.86}^{+0.85}$ Gyr.
 Furthermore, the halo appears to have a measurable age spread with standard deviation
 $1.18_{-0.62}^{+0.57}$ Gyr. 
 This result is preliminary as we await Gaia parallaxes to tightly constrain distances,
  which constrains both ages and stellar space motions.  
  If one or a few of these WDs have anomalous atmospheres, are unresolved binaries, 
  or are not true halo members, including them in this hierarchical analysis could artificially
   increase the estimated halo age spread.

Our mean halo age estimate is consistent with other WD-based age measurements for the Galactic halo.  For halo field WDs, these estimates are 11.4 $\pm$ 0.7 Gyr \citep{kalirai2012age}, 11-11.5 Gyr \citep{kilic2012},
 and $10.5^{+2.0}_{-1.5}$ \citep{kilic2017a}.
 Although broadly consistent, these studies all use somewhat 
 different techniques.  The study of \citet{kalirai2012age} relies on spectroscopic determinations of field and 
 globular cluster WDs.  The \citet{kilic2012} analysis depends on photometry and trigonometric parallaxes,
 as does our work, yet at that point only two halo WDs were available for their study. The \citet{kilic2017a} 
 study is based on the halo WD luminosity function isolated by \citet{munn2017a}.  Although this sample 
 contains 135 likely halo WDs, there are as yet no trigonometric parallaxes or spectroscopy to independently 
 constrain their masses.  Thus, all of these samples suffer some defects, and it is comforting to see that 
 different approaches to these different halo WD datasets yield consistent halo ages.

Another comparison to the field WD halo age is the WD age of those globular clusters that have halo 
properties.  Three globular clusters have been observed to sufficient depth to obtain their WD ages, and two of 
these (M4 and NGC 6397) exhibit halo kinematics and abundances.  The WD age of M4 is $11.6\pm2$ Gyr 
\citep{bedin2009end} and that age for NGC 6397 is $11.7\pm0.3$ Gyr \citep{hansen2013age}. 
These halo ages 
for globular cluster stars are almost identical to those for the halo field.  If there is any problem with these ages, 
it may only be that they are too young, at $\sim2$ Gyr younger than the age of the Universe.  At this point, we 
lack sufficient data to determine whether this is a simple statistical error, with most techniques having 
uncertainties in the range of $1$ Gyr, or whether there is a systematic error with the WD models or IFMR for 
these stars, or whether these WD studies have simply failed to find the oldest Galactic stars.  Alternatively, as 
mentioned above, the field halo age dispersion may really be of order $\pm2$ Gyr, in which case the halo field 
is sufficiently old, yet these globular clusters may not be.  We look forward to future results from Gaia and 
LSST that should reduce the observational errors in WD studies substantially while dramatically 
increasing sample size. 
This will allow us to precisely measure the age distribution of the Galactic halo and place
 the globular cluster ages into this context.

\section{Conclusion}
We propose the use of hierarchical modelling, fit via
EB and FB to obtain shrinkage estimates of the object-level parameters
of a population of objects.
We have developed novel computational algorithms to
fit hierarchical models even when the likelihood function is complicated. Our new algorithms
are able to take advantage of available case-by-case code, with substantial savings
in software development effort.  

By applying hierarchical modelling to a group of $10$ Galactic halo WDs, we estimate that
68.3\% of Galactic halo WDs have ages
 between $10.58$ and $13.51$ Gyr.
This tight age constraint from the photometry of only $10$ halo WDs 
demonstrates the power of our Bayesian hierarchical analysis.
In the near future, we expect not only better calibrated photometry for
many more WDs, but also to incorporate highly informative priors on distance and
population membership from the Gaia satellite's exquisite astrometry.
We look forward to using theses WDs to fit out hierarchical model
in order to derive an accurate and precise Galactic halo age distribution.

\section*{Acknowledgements}
We would like to thank the Imperial College High Performance Computing support team for their kind help, which greatly 
accelerates the simulation study in this project. Shijing Si thanks the Department of Mathematics in
Imperial College London for a Roth studentship, which supports his research. 
David van Dyk acknowledges support from a Wolfson Research Merit Award (WM110023) provided by 
the British Royal Society and from Marie-Curie Career Integration (FP7-PEOPLE-2012-CIG-321865) and 
Marie-Skodowska-Curie RISE (H2020-MSCA-RISE-2015-691164) Grants both provided by the European Commission. 


\bibliographystyle{mnras}
\bibliography{bibfile} 

\begin{thebibliography}{}
\expandafter\ifx\csname natexlab\endcsname\relax\def\natexlab#1{#1}\fi

\bibitem[{Barnes(2010)}]{barnes2010simple}
Barnes, S.~A. 2010, ApJ, 722, 222

\bibitem[{Bedin {et~al.}(2009)Bedin, Salaris, Piotto, Anderson, King, \&
  Cassisi}]{bedin2009end}
Bedin, L.~R., Salaris, M., Piotto, G., {et~al.} 2009, ApJ, 697, 965

\bibitem[{{Bergeron} {et~al.}(1995){Bergeron}, {Wesemael}, \&
  {Beauchamp}}]{bergeron1995}
{Bergeron}, P., {Wesemael}, F., \& {Beauchamp}, A. 1995, PASP, 1047

\bibitem[{{Carlin} \& {Louis}(2000)}]{carlin2000}
{Carlin}, B.~P., \& {Louis}, T.~A. 2000, Journal of the American Statistical
  Association, 95, 1286

\bibitem[{Carrasco {et~al.}(2014)Carrasco, Catalan, Jordi, Tremblay,
  Napiwotzki, Luri, Robin, \& Kowalski}]{carrasco2014gaia}
Carrasco, J., Catalan, S., Jordi, C., {et~al.} 2014, A\&A, 565, A11

\bibitem[{Casella(1985)}]{casella1985introduction}
Casella, G. 1985, The American Statistician, 39, 83

\bibitem[{{Clementini} {et~al.}(2003){Clementini}, {Gratton}, {Bragaglia},
  {Carretta}, {Di Fabrizio}, \& {Maio}}]{clementini2003}
{Clementini}, G., {Gratton}, R., {Bragaglia}, A., {et~al.} 2003, AJ, 125, 1309

\bibitem[{Dame {et~al.}(2016)Dame, Gianninas, Kilic, Munn, Brown, Williams, von
  Hippel, \& Harris}]{dame2016new}
Dame, K., Gianninas, A., Kilic, M., {et~al.} 2016, MNRAS, 463, 2453

\bibitem[{DeGennaro {et~al.}(2009)DeGennaro, von Hippel, Jefferys, Stein, van
  Dyk, \& Jeffery}]{degennaro2009inverting}
DeGennaro, S., von Hippel, T., Jefferys, W.~H., {et~al.} 2009, ApJ, 696, 12

\bibitem[{{Dempster} {et~al.}(1977){Dempster}, {Laird}, \&
  {Rubin}}]{dempster1977}
{Dempster}, A.~P., {Laird}, N.~M., \& {Rubin}, D.~B. 1977, Journal of the royal
  statistical society. Series B (methodological), 1

\bibitem[{Di~Benedetto(1997)}]{di1997improved}
Di~Benedetto, G. 1997, ApJ, 486, 60

\bibitem[{{Dotter} {et~al.}(2008){Dotter}, {Chaboyer}, {Jevremovi{\'c}},
  {Kostov}, {Baron}, \& {Ferguson}}]{dotter2008}
{Dotter}, A., {Chaboyer}, B., {Jevremovi{\'c}}, D., {et~al.} 2008, ApJ
  Supplement Series, 178, 89

\bibitem[{{Efron}(1979)}]{efron1979}
{Efron}, B. 1979, The annals of Statistics, 1

\bibitem[{{Efron}(1996)}]{efron1996}
---. 1996, Journal of the American Statistical Association, 91, 538

\bibitem[{{Efron} \& {Morris}(1972)}]{efron1972}
{Efron}, B., \& {Morris}, C. 1972, Journal of the American Statistical
  Association, 67, 130

\bibitem[{Eggen {et~al.}(1962)Eggen, Lynden-Bell, \&
  Sandage}]{eggen1962evidence}
Eggen, O.~J., Lynden-Bell, D., \& Sandage, A.~R. 1962, ApJ, 136, 748

\bibitem[{Feast(2000)}]{feast2000local}
Feast, M. 2000, https://arxiv.org/pdf/astro-ph/0010590.pdf

\bibitem[{Feast \& Catchpole(1997)}]{feast1997the}
Feast, M.~W., \& Catchpole, R.~M. 1997, MNRAS, 286, L1

\bibitem[{Fitzpatrick {et~al.}(2002)Fitzpatrick, Ribas, Guinan, DeWarf,
  Maloney, \& Massa}]{fitzpatrick2002fundamental}
Fitzpatrick, E.~L., Ribas, I., Guinan, E.~F., {et~al.} 2002, ApJ, 564, 260

\bibitem[{Fontaine {et~al.}(2001)Fontaine, Brassard, \&
  Bergeron}]{fontaine2001potential}
Fontaine, G., Brassard, P., \& Bergeron, P. 2001, PASP, 113, 409

\bibitem[{Forbes \& Bridges(2010)}]{forbes2010accreted}
Forbes, D.~A., \& Bridges, T. 2010, MNRAS, 404, 1203

\bibitem[{Freeman \& Bland-Hawthorn(2002)}]{freeman2002new}
Freeman, K., \& Bland-Hawthorn, J. 2002, ARAA, 40, 487

\bibitem[{{Gelman}(2006)}]{gelman2006}
{Gelman}, A. 2006, Technometrics, 48, 241

\bibitem[{{Gelman} {et~al.}(2014){Gelman}, {Carlin}, {Stern}, \&
  {Rubin}}]{gelman2014}
{Gelman}, A., {Carlin}, J.~B., {Stern}, H.~S., \& {Rubin}, D.~B. 2014, Bayesian
  data analysis, Vol.~2 (Taylor \& Francis)

\bibitem[{Gelman {et~al.}(2006)}]{gelman2006prior}
Gelman, A., {et~al.} 2006, Bayesian analysis, 1, 515

\bibitem[{Gianninas {et~al.}(2015)Gianninas, Curd, Thorstensen, Kilic,
  Bergeron, Andrews, Canton, \& Ag{\"u}eros}]{gianninas2015ultracool}
Gianninas, A., Curd, B., Thorstensen, J.~R., {et~al.} 2015, MNRAS, 449, 3966

\bibitem[{Gieren {et~al.}(1998)Gieren, Fouqu{\'e}, \&
  G{\'o}mez}]{gieren1998cepheid}
Gieren, W.~P., Fouqu{\'e}, P., \& G{\'o}mez, M. 1998, ApJ, 496, 17

\bibitem[{Gratton {et~al.}(2004)Gratton, Sneden, \&
  Carretta}]{gratton2004abundance}
Gratton, R., Sneden, C., \& Carretta, E. 2004, Annu. Rev. Astron. Astrophys.,
  42, 385

\bibitem[{Groenewegen(2000)}]{groenewegen2000lmc}
Groenewegen, M.~A. 2000, arXiv preprint astro-ph/0010298

\bibitem[{Hansen {et~al.}(2013)Hansen, Kalirai, Anderson, Dotter, Richer, Rich,
  Shara, Fahlman, Hurley, King, {et~al.}}]{hansen2013age}
Hansen, B., Kalirai, J.~S., Anderson, J., {et~al.} 2013, Nature, 500, 51

\bibitem[{Hills {et~al.}(2015)Hills, von Hippel, Courteau, \&
  Geller}]{hills2015bayesian}
Hills, S., von Hippel, T., Courteau, S., \& Geller, A.~M. 2015, AJ, 149, 94

\bibitem[{James \& Stein(1961)}]{james1961}
James, W., \& Stein, C. 1961, in Proceedings of the fourth Berkeley symposium
  on mathematical statistics and probability, Vol.~1, 361--379

\bibitem[{Jeffery {et~al.}(2011)Jeffery, von Hippel, DeGennaro, van Dyk, Stein,
  \& Jefferys}]{jeffery2011white}
Jeffery, E.~J., von Hippel, T., DeGennaro, S., {et~al.} 2011, ApJ, 730, 35

\bibitem[{Kalirai(2012)}]{kalirai2012age}
Kalirai, J.~S. 2012, Nature, 486, 90

\bibitem[{Kilic {et~al.}(2017)Kilic, Munn, Harris, von Hippel, Liebert,
  Williams, Jeffery, \& S}]{kilic2017a}
Kilic, M., Munn, J.~A., Harris, H.~C., {et~al.} 2017, submitted to AJ

\bibitem[{{Kilic} {et~al.}(2012){Kilic}, {Thorstensen}, {Kowalski}, \&
  {Andrews}}]{kilic2012}
{Kilic}, M., {Thorstensen}, J.~R., {Kowalski}, P., \& {Andrews}, J. 2012,
  MNRAS, 423, L132

\bibitem[{{Kilic} {et~al.}(2010){Kilic}, {Munn}, {Williams}, {Kowalski}, {von
  Hippel}, {Harris}, {Jeffery}, {DeGennaro}, {Brown}, \& {McLeod}}]{kilic2010}
{Kilic}, M., {Munn}, J.~A., {Williams}, K.~A., {et~al.} 2010, ApJ, 715, L21

\bibitem[{Laney \& Stobie(1994)}]{laney1994cepheid}
Laney, C., \& Stobie, R. 1994, MNRAS, 266, 441

\bibitem[{Leaman {et~al.}(2013)Leaman, VandenBerg, \&
  Mendel}]{leaman2013bifurcated}
Leaman, R., VandenBerg, D.~A., \& Mendel, J.~T. 2013, MNRAS, stt1540

\bibitem[{Licquia \& Newman(2015)}]{licquia2015improved}
Licquia, T.~C., \& Newman, J.~A. 2015, ApJ, 806, 96

\bibitem[{Mandel {et~al.}(2009)Mandel, Wood-Vasey, Friedman, \&
  Kirshner}]{mandel2009type}
Mandel, K.~S., Wood-Vasey, W.~M., Friedman, A.~S., \& Kirshner, R.~P. 2009,
  ApJ, 704, 629

\bibitem[{March {et~al.}(2014)March, Karpenka, Feroz, \&
  Hobson}]{march2014comparison}
March, M., Karpenka, N.~V., Feroz, F., \& Hobson, M. 2014, MNRAS, 437, 3298

\bibitem[{{Miller} \& {Scalo}(1979)}]{miller1979}
{Miller}, G.~E., \& {Scalo}, J.~M. 1979, ApJ Supplement Series, 41, 513

\bibitem[{Montgomery {et~al.}(1999)Montgomery, Klumpe, Winget, \&
  Wood}]{montgomery1999evolutionary}
Montgomery, M., Klumpe, E., Winget, D., \& Wood, M. 1999, ApJ, 525, 482

\bibitem[{{Morris}(1983)}]{morris1983}
{Morris}, C.~N. 1983, Journal of the American Statistical Association, 78, 47

\bibitem[{Morris \& Lysy(2012)}]{morris2012shrinkage}
Morris, C.~N., \& Lysy, M. 2012, Statistical Science, 115

\bibitem[{Munn {et~al.}(2017)Munn, Harris, von Hippel, Kilic, Liebert,
  Williams, DeGennaro, \& et~al.}]{munn2017a}
Munn, J.~A., Harris, H.~C., von Hippel, T., {et~al.} 2017, in press

\bibitem[{{O'Malley} {et~al.}(2013){O'Malley}, {von Hippel}, \& {van
  Dyk}}]{o2013}
{O'Malley}, E.~M., {von Hippel}, T., \& {van Dyk}, D.~A. 2013, ApJ, 775, 1

\bibitem[{Panagia(1998)}]{panagia1998new}
Panagia, N. 1998, Memorie della Societa Astronomica Italiana, 69, 225

\bibitem[{Park {et~al.}(2008)Park, van Dyk, \&
  Siemiginowska}]{park2008searching}
Park, T., van Dyk, D.~A., \& Siemiginowska, A. 2008, ApJ, 688, 807

\bibitem[{Pawlowski {et~al.}(2012)Pawlowski, Pflamm-Altenburg, \&
  Kroupa}]{pawlowski2012vpos}
Pawlowski, M., Pflamm-Altenburg, J., \& Kroupa, P. 2012, MNRAS, 423, 1109

\bibitem[{Pietrzy{\'n}ski \& Gieren(2002)}]{pietrzynski2002araucaria}
Pietrzy{\'n}ski, G., \& Gieren, W. 2002, AJ, 124, 2633

\bibitem[{Renedo {et~al.}(2010)Renedo, Althaus, Bertolami, Romero, C{\'o}rsico,
  Rohrmann, \& Garcia-Berro}]{renedo2010new}
Renedo, I., Althaus, L.~G., Bertolami, M.~M., {et~al.} 2010, ApJ, 717, 183

\bibitem[{Roederer {et~al.}(2010)Roederer, Sneden, Thompson, Preston, \&
  Shectman}]{roederer2010characterizing}
Roederer, I.~U., Sneden, C., Thompson, I.~B., Preston, G.~W., \& Shectman,
  S.~A. 2010, ApJ, 711, 573

\bibitem[{Romaniello {et~al.}(2000)Romaniello, Salaris, Cassisi, \&
  Panagia}]{romaniello2000hubble}
Romaniello, M., Salaris, M., Cassisi, S., \& Panagia, N. 2000, ApJ, 530, 738

\bibitem[{Sarajedini {et~al.}(2002)Sarajedini, Grocholski, Levine, \&
  Lada}]{sarajedini2002k}
Sarajedini, A., Grocholski, A.~J., Levine, J., \& Lada, E. 2002, AJ, 124, 2625

\bibitem[{Scannapieco {et~al.}(2011)Scannapieco, White, Springel, \&
  Tissera}]{scannapieco2011formation}
Scannapieco, C., White, S.~D., Springel, V., \& Tissera, P.~B. 2011, MNRAS,
  417, 154

\bibitem[{Shariff {et~al.}(2016)Shariff, Dhawan, Jiao, Leibundgut, Trotta, \&
  van Dyk}]{shariff2016standardizing}
Shariff, H., Dhawan, S., Jiao, X., {et~al.} 2016, MNRAS, 463, 4311

\bibitem[{Si {et~al.}(2017)Si, van Dyk, von Hippel, \& Robinson}]{si2017}
Si, S., van Dyk, D., von Hippel, T., \& Robinson, E. 2017, in preparation

\bibitem[{Soderblom(2010)}]{soderblom2010ages}
Soderblom, D.~R. 2010, arXiv preprint arXiv:1003.6074

\bibitem[{{Stein} {et~al.}(2013){Stein}, {van Dyk}, {von Hippel}, {DeGennaro},
  {Jeffery}, \& {Jefferys}}]{stein2013}
{Stein}, N.~M., {van Dyk}, D.~A., {von Hippel}, T., {et~al.} 2013, Statistical
  Analysis and Data Mining: The ASA Data Science Journal, 6, 34

\bibitem[{Tumlinson {et~al.}(2010)Tumlinson, Malec, Carswell, Murphy, Buning,
  Milutinovic, Ellison, Prochaska, Jorgenson, Ubachs,
  {et~al.}}]{tumlinson2010cosmological}
Tumlinson, J., Malec, A., Carswell, R., {et~al.} 2010, ApJ, 718, L156

\bibitem[{{van Dyk} {et~al.}(2001){van Dyk}, {Connors}, {Kashyap}, \&
  {Siemiginowska}}]{van2001analysis}
{van Dyk}, D.~A., {Connors}, A., {Kashyap}, V.~L., \& {Siemiginowska}, A. 2001,
  \apj, 548, 224

\bibitem[{{van Dyk} {et~al.}(2009){van Dyk}, {Degennaro}, {Stein}, {Jefferys},
  \& {von Hippel}}]{van2009}
{van Dyk}, D.~A., {Degennaro}, S., {Stein}, N., {Jefferys}, W.~H., \& {von
  Hippel}, T. 2009, Annals of Applied Statistics, 3, 117

\bibitem[{van Dyk \& Meng(2010)}]{van2010cross}
van Dyk, D.~A., \& Meng, X.-L. 2010, Statist. Sci., 25, 429

\bibitem[{van Leeuwen {et~al.}(1997)van Leeuwen, Feast, Whitelock, \&
  Yudin}]{van1997first}
van Leeuwen, F., Feast, M., Whitelock, P., \& Yudin, B. 1997, MNRAS, 287, 955

\bibitem[{{von Hippel} {et~al.}(2006){von Hippel}, {Jefferys}, {Scott},
  {Stein}, {Winget}, {De Gennaro}, {Dam}, \& {Jeffery}}]{von2006inverting}
{von Hippel}, T., {Jefferys}, W.~H., {Scott}, J., {et~al.} 2006, \apj, 645,
  1436

\bibitem[{{Wei} \& {Tanner}(1990)}]{wei1990}
{Wei}, G.~C., \& {Tanner}, M.~A. 1990, Journal of the American statistical
  Association, 85, 699

\bibitem[{{Williams} {et~al.}(2009){Williams}, {Bolte}, \&
  {Koester}}]{williams2009}
{Williams}, K.~A., {Bolte}, M., \& {Koester}, D. 2009, ApJ, 693, 355

\bibitem[{Yamada {et~al.}(2013)Yamada, Suda, Komiya, Aoki, \&
  Fujimoto}]{yamada2013stellar}
Yamada, S., Suda, T., Komiya, Y., Aoki, W., \& Fujimoto, M.~Y. 2013, MNRAS,
  436, 1362

\end{thebibliography}


\appendix

\section{Shrinkage Estimates}\label{sec:shrinkage}

In this appendix, we discuss shrinkage estimates and their advantages.
Consider, for example, a Gaussian model,
\begin{equation}\label{eq:norm}
{Y}_{i}|\theta_{i}\sim{N}(\theta_{i},\sigma),~i=1, 2, \cdots, n,
\end{equation}
where $\bm{Y}=({Y}_{1},\ldots,{Y}_{n})$ is a vector
of independent observations of each of $n$ objects,
 $\bm{\theta}=(\theta_{1},\ldots, \theta_{n})$ is the vector of object-level parameters of interest,
 and $\sigma$
  is the known measurement error.
 A simple technique to fit Eq. \ref{eq:norm} is to estimate each $\theta_{i}$ individually, 
  $\hat{\theta}^{\rm ind}_{i}={Y}_{i}$, i.e. $\hat{\bm{\theta}}^{\rm ind}=\bm{Y}$,
  using only its corresponding data.
   If a population is believed to be homogeneous, however, we might suppose, 
   in the extreme case, that all of the objects
  have the same parameter, $\theta_{1}=\theta_{2}=\cdots=\theta_{n}$. For example, one might suppose
  stars in a cluster all have the same age. Under this assumption, the
  pooled estimate, $\hat{\theta}_{i}^{\rm pool}=\bar{Y}=\frac{1}{n}\sum{Y}_{i}$,
  i.e. $\hat{\bm{\theta}}^{\rm pool}=(\bar{Y},\ldots,\bar{Y})$, is appropriate.
  
  The mean squared error (MSE) is a statistical quantity that can be used to evaluate the quality of
  an estimator. As its name implies, it measures the average of the squared deviation between
  the estimator and true parameter value. Thus the MSE of $\hat{\bm{\theta}}^{\rm ind}$ is
$${\rm MSE}(\hat{\bm{\theta}}^{\rm ind})={\rm E}\bigg[\sum_{i=1}^{n}(\hat{\theta}^{\rm ind}_{i}-\theta_{i})^{2}|\bm{\theta}\bigg]=n\sigma^2,$$
where ${\rm E}(\cdot|\bm{\theta})$ is the conditional expectation function that
assumes $\bm{\theta}$ is fixed and here that $\hat{\bm{\theta}}^{\rm ind}=\bm{Y}$ varies according to
the model in Eq. \ref{eq:norm}.
It is well known in the statistics literature that the individual estimators
$\hat{\bm{\theta}}^{\rm ind}$ are \textit{inadmissible}
if $n>3$. This means that there is another estimator that has smaller MSE regardless of the true values of
$\bm{\theta}$ or $\sigma^{2}$. In particular, the James-Stein estimator 
 of $\bm{\theta}$,
  $\hat{\bm{\theta}}^{\rm JS}=(1-\hat{B})\hat{\bm{\theta}}^{\rm ind}+\hat{B}\hat{\bm{\theta}}^{\rm pool}$ 
  where
 ${S}^2=\sum(Y_{i}-\bar{Y})^{2}/(n-1)$ and $\hat{B}=(n-3)\sigma^2/(n-1){S}^{2}$,
 is known to have smaller MSE than $\hat{\bm{\theta}}^{\rm ind}$
 if $n >3$ \citep{james1961, efron1972, morris1983}\footnote{ It can be shown that 
 the MSE of $\hat{\bm{\theta}}^{\rm JS}$ is 
 \begin{eqnarray*}
 {\rm E}\bigg[\sum_{i=1}^{n}(\hat{\theta}^{\rm JS}_{i}-\theta_{i})^{2}|\bm{\theta}\bigg]&=&n\sigma^{2}-\sigma^{2}(n-3){\rm E}(\hat{B})\\
 &<&n\sigma^{2}={\rm E}\bigg[\sum_{i=1}^{n}(\hat{\theta}^{\rm ind}_{i}-\theta_{i})^{2}|\bm{\theta}\bigg],
 \end{eqnarray*}
 which shows the advantage of James-Stein estimator in terms of MSE over the individual estimator when $n>3$.}.
 When $n >3$, $\hat{B}>0$ and the James-Stein estimator of $\theta_{i}$ is a weighted average of the individual
 estimates, $\hat{\theta}_{i}^{\rm ind}=Y_{i}$, and the pooled estimates,
  $\hat{\theta}^{\rm pool}_{i}=\bar{Y}$.The James-Stein estimator is an example of a {\it shrinkage estimators}, 
 which are estimates of a set of object-level parameters that are ``shrunk'' toward a common 
 central value relative to those derived from the corresponding case-by-case analyses. 

The population-level parameters that describe the distribution of $(\theta_{1}, \ldots, \theta_{n})$ are 
often also of interest. Suppose
 we model the population by assuming that $\theta_{i}$ follows a common normal distribution, i.e., 
 we extend the model in Eq. \ref{eq:norm} to
 \begin{eqnarray}
{Y}_{i}|\theta_{i}&\sim&{N}(\theta_{i},\sigma),~i=1,2,\cdots,n;\label{eq:ex1}\\
\theta_{i}&\sim&{N}(\gamma,\tau^{2}),\label{eq:ex2}
\end{eqnarray}
where $\gamma$ and $\tau$ are unknown population-level parameters. The model in 
Eqs.~\ref{eq:ex1}-\ref{eq:ex2} is a hierarchical model and can be fit using
 Empirical Bayes (EB) \citep[e.g.][]{morris1983,efron1996}.
 We choose the non-informative, ${p}(\gamma, \tau)\propto 1$, which is
a standard choice in this setting \citep[e.g.,][]{gelman2006prior}.
 The EB approach is Bayesian
 in that it views Eq. \ref{eq:ex2} as a prior distribution and is empirical in that the parameters of
 this prior are fit to the data. Specifically, 
 EB proceeds by first deriving the marginal posterior
 distribution of $\gamma$ and $\tau^{2}$, 
 \begin{equation}\label{eq:eb_ex}
 {p}(\gamma,\tau^{2}|\bm{Y})={p}(\gamma, \tau^{2})\prod_{i=1}^{n}\int{p}(Y_{i}|\theta_{i}){p}(\theta_{i}|\gamma,\tau^{2}){\rm d}\theta_{i},
 \end{equation}
and then estimating $\gamma$ and $\tau^{2}$ with the values that maximise Eq. \ref{eq:eb_ex}. 
These estimates are
 $\hat{\gamma}=\bar{Y}$ and $\hat{\tau}^{2}=\max\bigg\{\frac{\sum_{i=1}^{n}(Y_{i}-\bar{Y})^{2}}{n+1}-\sigma^{2},0\bigg\}$.
(Even with the normal assumptions in Eq. \ref{eq:ex1}-\ref{eq:ex2}, closed form estimates of $\gamma$ and $\tau^{2}$ are available only under the simplifying assumption
that the measurement errors for each $Y_{i}$ are the same, i.e., $\sigma^{2}_{1}=\sigma_{2}^{2}=\cdots=\sigma_{n}^{2}$.)
 Finally, the posterior distribution of $\theta_{i}$ can be expressed as
 \begin{equation}\label{eq:eb_ex2}
{p}(\theta_{i}|Y_{i},\hat{\gamma}, \hat{\tau}^{2})\propto{p}(Y_{i}|\theta_{i}){p}(\theta_{i}|\hat{\gamma},\hat{\tau}^{2}).
\end{equation}
 Each $\theta_{i}$ can be estimated with its posterior mean under Eq. \ref{eq:eb_ex2}.
Under certain conditions, EB is consistent with James-Stein estimators \citep[e.g.][]{morris1983}.
EB can produce estimators having the same advantages as James-Stein and it is readily able
 to handle more complicated problems whereas James-Stein would require model specific derivation
 of MSE-reducing estimators.
 
\section{Large Magellanic Cloud}\label{sec:lmc}

We illustrate the construction and fitting of hierarchical models 
and the advantages of shrinkage estimates through
an illustrative application to data used to estimate the distance to the LMC.
The LMC is a 
satellite galaxy of the Milky Way.  
Numerous estimates based on various data sources have been made of
 the distance modulus to the LMC. The population of stars used affects
 the estimated distance modulus: Estimates based on Population I tend to be larger than
 those based on Population II stars.
We use a set of estimates based on Population I stars, and formulate a hierarchical model
 for these estimates in order to develop a comprehensive estimate.
We use the data in Table \ref{table:data}, which was compiled by \citet{clementini2003}.
\begin{table*}
\caption{Population I Distance Indicators\label{table:data}}
\centering
\begin{tabular}{lcl}
\hline\hline
Method & Reported Distance Modulus & References \\ \hline
Cepheids: trig. paral. & 18.70 $\pm$ 0.16 & \citet{feast1997the} \\
Cepheids: MS fitting & 18.55 $\pm$ 0.06 & \citet{laney1994cepheid} \\
Cepheids: B-W & 18.55 $\pm$ 0.10 & \citet{gieren1998cepheid, di1997improved} \\
Cepheids: P/L relation & 18.575 $\pm$ 0.2 & \citet{groenewegen2000lmc} \\
Eclipsing binaries & 18.4 $\pm$ 0.1 & \citet{fitzpatrick2002fundamental} \\
Clump & 18.42 $\pm$ 0.07 & \citet{clementini2003} \\
Clump & 18.45 $\pm$ 0.07 & \citet{clementini2003} \\
Clump & 18.59 $\pm$ 0.09 & \citet{romaniello2000hubble} \\
Clump & 18.471 $\pm$ 0.12 & \citet{pietrzynski2002araucaria} \\
Clump & 18.54 $\pm$ 0.10 & \citet{sarajedini2002k} \\
Miras & 18.54 $\pm$ 0.18 & \citet{van1997first} \\
Miras & 18.54 $\pm$ 0.14 & \citet{feast2000local} \\
SN 1987a & 18.54 $\pm$ 0.05 & \citet{panagia1998new}\\ 
\hline
\end{tabular}
\end{table*}

Besides statistical errors, the various distance estimates may be subject to systematic errors.
We aim to estimate the magnitude of
 these systematic errors. If we further assume that the systematic errors
tend to average out among the various estimators,
we can obtain a better comprehensive estimator of the distance modulus. 
Let $\mu_{i}$ be the best estimate of the distance modulus that could be
obtained with method $i$, i.e., with an arbitrarily large dataset. Because of systematic
errors, $\mu_{i}$ does not equal the true distance modulus, but is free of statistical error.

Consider the statistical model, 
\begin{eqnarray}
{D}_{i}&\sim&{N}(\mu_{i},\sigma_{i}),~i=1,\cdots,13, \label{eq:level1}  \\
\mu_{i}&\sim&{N}(\gamma,\tau), \label{eq:level2}
\end{eqnarray}
where
${D}_{i}$ is the actual estimated distance modulus based on the method/dataset $i$ including statistical error, $\sigma_{i}$ is
the known standard deviation of the statistical error,
 $\gamma$ is the true distance modulus of the LMC, and $\tau$ is the standard deviation 
 of the systematic errors of the various estimates. Eq. \ref{eq:level2} specifies our assumption
 that the systematic errors tend to average out. 
We denote $\bm{D}=({D}_{1},\cdots,{D}_{13})$ and $\bm{\mu}=(\mu_{1}, \cdots, \mu_{13})$.
 
We take an EB approach to fitting the hierarchical model in Eq.~\ref{eq:level1}--\ref{eq:level2}.
This involves first estimating the population-level parameters $\gamma$ and $\tau$ and then plugging these
 estimates in
Eq. \ref{eq:level2} and using it as the prior distribution for each $\mu_{i}$.
Finally the individual $\mu_{i}$ are estimated with their posterior expectations, 
${\rm E}(\mu_{i}|\bm{D}, \hat{\gamma}, \hat{\tau})$ and their posterior standard deviations,
$SD(\mu_{i}|\bm{D}, \hat{\gamma}, \hat{\tau})$ are used as $1\sigma$ uncertainties.
 Our EB approach requires a prior distribution for $\gamma$
and $\tau$. We choose the standard non-informative prior,
${p}(\gamma, \tau)\propto 1$ in this setting. 

We estimate $\gamma$ and $\tau$ by maximising their joint posterior density, 
\begin{eqnarray}
{p}(\gamma,\tau|\bm{D})&\propto&\int{p}(\tau,\gamma,\bm{\mu}|\bm{D}){\rm d}\bm{\mu}\nonumber \\
&=&{p}(\gamma, \tau)\prod_{i=1}^{13}\int{p}({D}_{i}|\mu_{i}){p}(\mu_{i}|\gamma,\tau){\rm d}\mu_{i}.\label{eq:lmcjpost}
\end{eqnarray}
The values of $\gamma$ and $\tau$ that maximise Eq. \ref{eq:lmcjpost} are known as maximum a posterior (MAP)
estimates.
For any $\tau$, Eq.~\ref{eq:lmcjpost} is maximised with respect to $\gamma$ by 
\begin{equation}\label{eq:gamma}
\hat{\gamma}(\tau)=\frac{\sum_{i=1}^{13}{D}_{i}/(\tau^2+\sigma_{i}^{2})}
{\sum_{i=1}^{13}1/(\tau^2+\sigma_{i}^{2})},
\end{equation}
where $\hat{\gamma}(\tau)$ is a function of $\tau$.
The profile posterior density of $\tau$ is obtained by evaluating Eq.~\ref{eq:lmcjpost} at
$\hat{\gamma}(\tau)$ and $\tau$, i.e., ${p}(\hat{\gamma}(\tau),\tau|\bm{D})$.
The global maximiser of the profile posterior distribution is the MAP estimate of $\tau$
and must be obtained numerically.

\begin{figure*}
\centering
\includegraphics[width=0.8\textwidth]{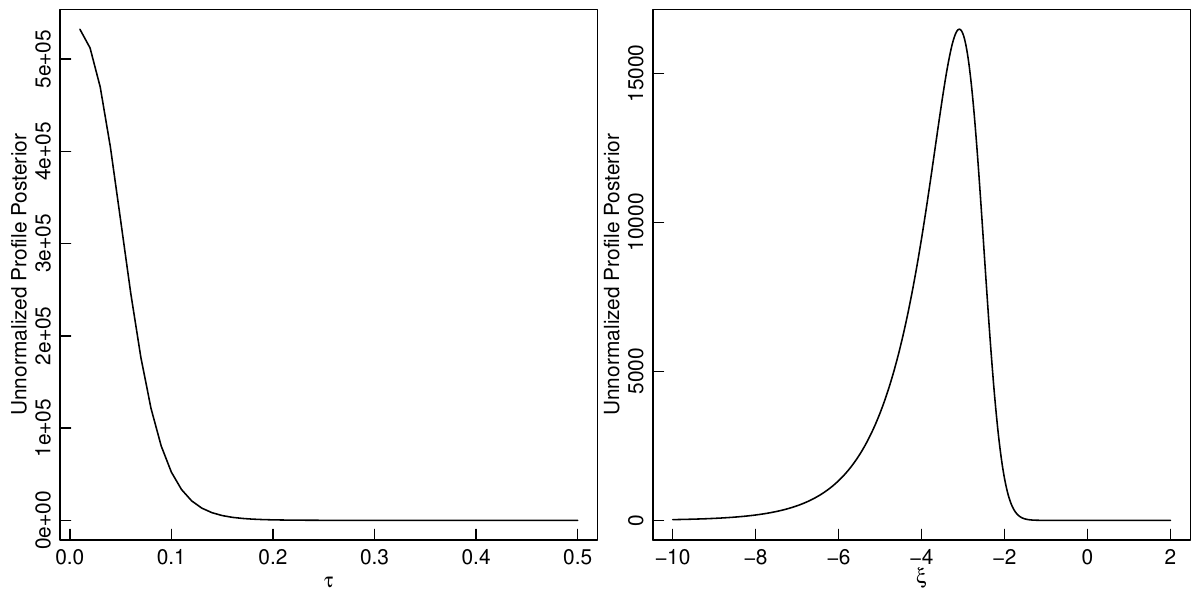}
\caption{The profile posterior distribution of $\tau$ (left panel) and of $\xi=\ln\tau$ (right panel). The modal 
estimate of $\tau$ is zero and is less representative of the distribution than is the modal estimate of $\xi$.
\label{fig:profile}}
\end{figure*}

As shown in the left panel of Fig.~\ref{fig:profile}, the profile posterior density of $\tau$ 
 monotonically decreases from its peak at $0$,
 which means that the MAP estimator of $\tau$ is $0$, a poor summary of the profile posterior.
 This is because $0$ is the lower boundary of the possible values of $\tau$.
 A better estimate can be obtained using a transformation of the population standard deviation,
 specifically, $\xi=\ln\tau$. The joint posterior of $\gamma$ and $\xi$ can be expressed as
$${p}(\gamma,\xi|\bm{D})=
{p}(\gamma,\exp(\xi)|\bm{D})\exp(\xi),$$
where ${p}(\cdot| \bm{D})$ is the posterior distribution of $\gamma$ and $\tau$.
The profile posterior of $\xi$ is plotted in
the right panel of Fig.~\ref{fig:profile}, is more symmetric, and is better summarised
by its mode. After having estimated $\xi$ with its MAP estimate, we compute
$\hat{\tau}=\exp(\hat{\xi})$ and $\hat{\gamma}=\gamma(\hat{\tau})$.
See \citet{park2008searching} for a discussion of transforming parameters to
achieve approximate symmetry in the case of mode-based estimates.

\begin{figure}
\centering
\includegraphics[width=.5\textwidth]{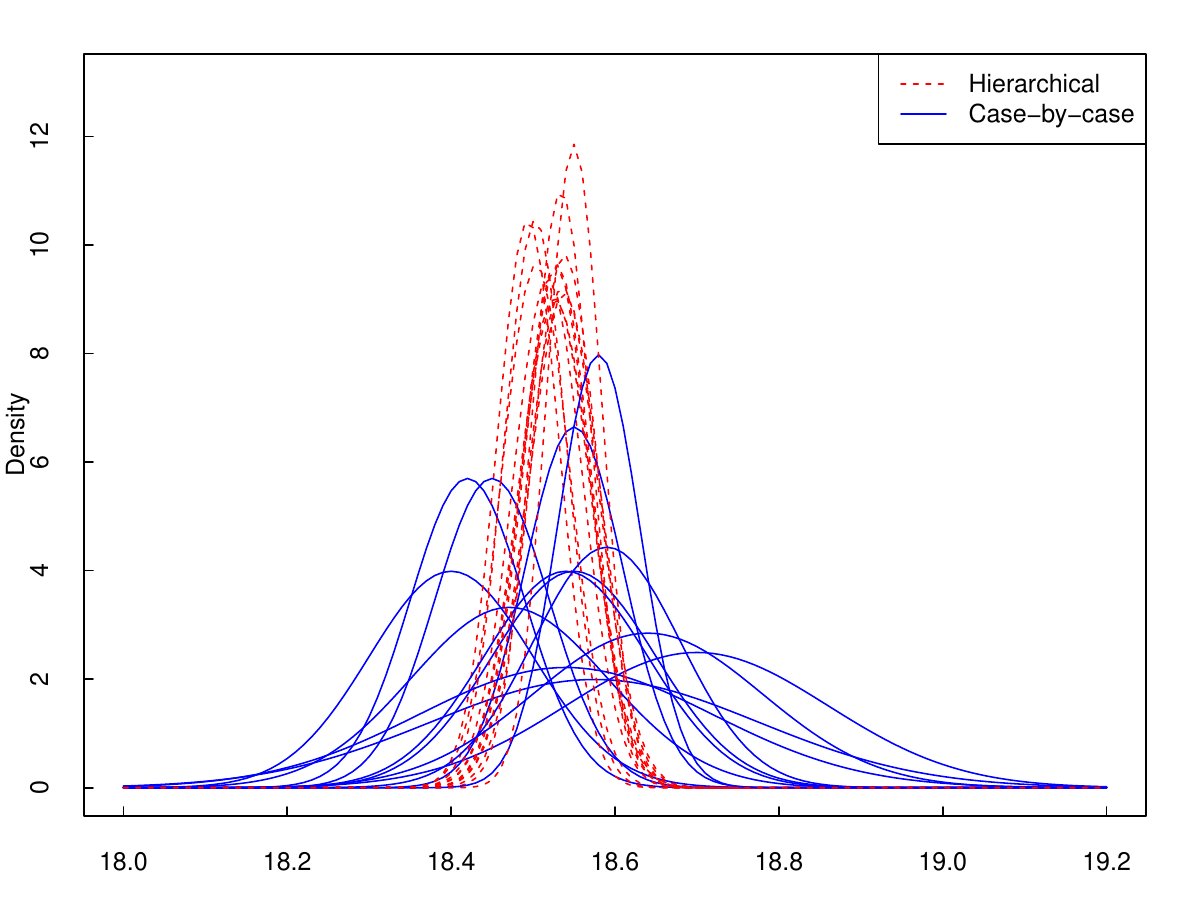}
\caption{Posterior density distributions of the LMC distance modulus based on hierarchical fitting (red dashed lines) and 
based on the case-by-case analysis (blue solid lines). Those based on hierarchical fitting shrink
towards the centre relative to those based on the case-by-case analysis.
\label{fig:distance}}
\end{figure}

Plugging $\hat{\gamma}$ and $\hat{\tau}$ into the prior for $\mu_{i}$ given in Eq. \ref{eq:level2}, we
can compute the posterior distribution of each $\mu_{i}$ as
$$
 {p}(\bm{\mu}|\bm{D}, \hat{\gamma}, \hat{\tau}^{2})\propto\prod_{i=1}^{13}{p}({D}_{i}|\mu_{i}){p}(\mu_{i}|\hat{\gamma},\hat{\tau}).$$
Fig.~\ref{fig:distance} shows the hierarchical and case-by-case posterior distributions of
the individual estimates, $\mu_{i}$. The hierarchical results (dashed red lines) are shrunk
toward the centre relative to the case-by-case results (blue solid lines).
The case-by-case density functions of $\mu_{i}$ range from $18.0$ to $19.2$, whereas
the hierarchical posterior density functions are more precise, ranging from $18.3$ to $18.7$.
  This is an example of the shrinkage of the case-by-case fits towards their average that occurs 
when fitting a hierarchical model. We can also see this effect in the posterior means,
\begin{equation}\label{eq:lmc.postmean}
{\rm E}(\mu_{i}|\tau,\bm{D})=\frac{\hat{\gamma}(\tau)/\tau^2+{D}_{i}/\sigma_{i}^{2}}{1/\tau^2+1/\sigma_{i}^{2}},
\end{equation}
which are weighted averages of the case-by-case estimates, $D_{i}$, and the combined 
(MAP) estimate of the
distance modulus, given in Eq.~\ref{eq:gamma}. The MAP estimate of the
distance modulus is $\hat{\gamma}=18.525$ and
the standard deviation of the systematic errors is $\hat{\tau}=0.045$ and the distance modulus
is $\hat{\gamma}=\gamma(\hat{\tau})=18.525$. 
We compute $\hat{\tau}$ via the MAP estimate of $\xi$ as described above.
It measures the
extent of heterogeneity between $13$ different published results.
To compute the uncertainty of $\hat{\gamma}$, we generate $200$ bootstrap samples \citep{efron1979}
of $\bm{D}$ and
for each we compute the MAP estimate for $\gamma$, resulting in $200$ bootstrap
estimates of $\gamma$ with standard deviation $0.024$. Thus our estimate of $\gamma$,
the distance modulus
of LMC, is $18.53\pm0.024$, that is, $50.72\pm0.56$ kpc.

\begin{figure*}
\centering
\includegraphics[width=0.8\textwidth]{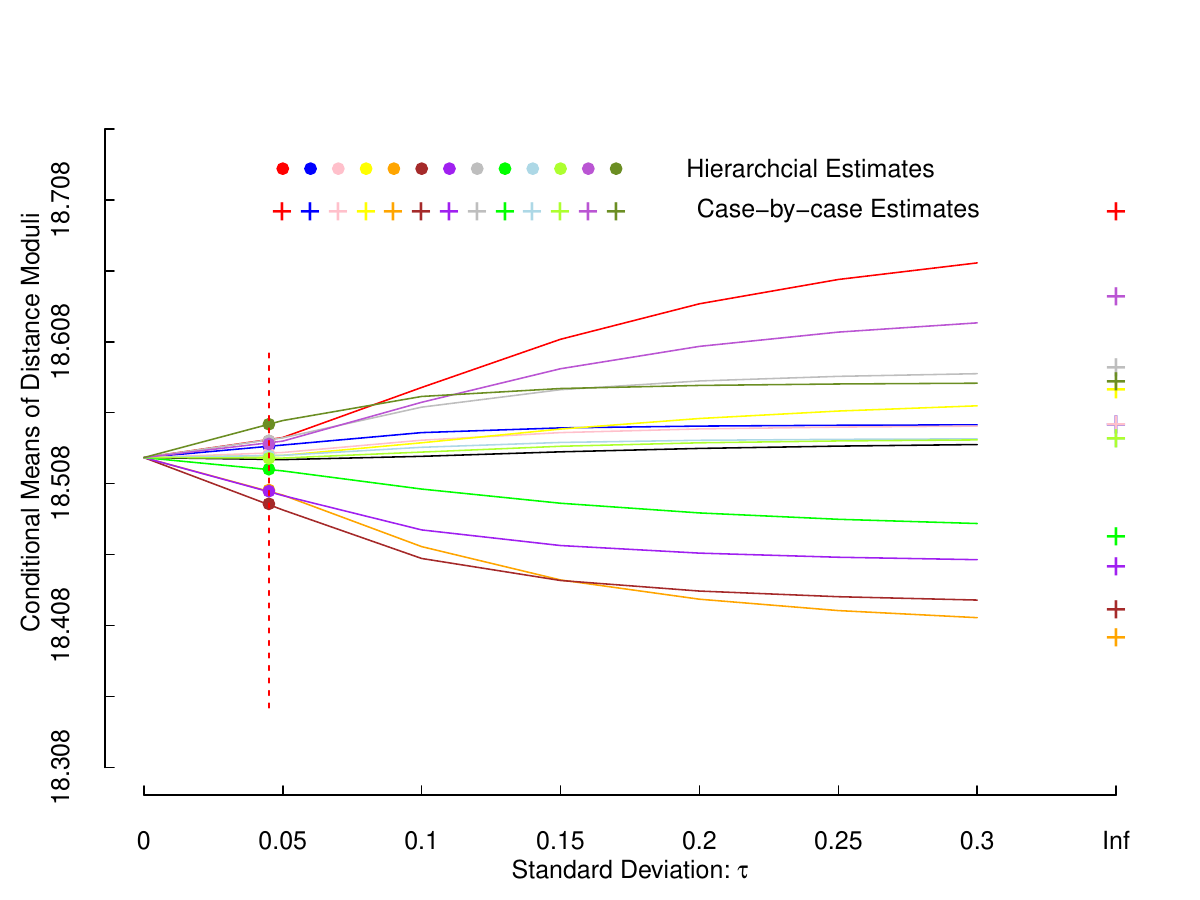}
\caption{Conditional posterior means of distance modulus ${\rm E}(\mu_{i}|\tau,\bm{D})$, as a function of
the standard deviation of the systematic errors $\tau$. The red dashed line indicates our estimate $\hat{\tau}=0.045$.
The 13 solid round dots of different colours show the hierarchical estimates of the distance modulus to the LMC using 
various methods, whereas the plus signs are estimates from the case-by-case analyses.
(Plus signs and round dots of the same colour correspond to the same published result; see Table~\ref{table:data}.) \label{fig:shrinkageplot}}
\end{figure*}

For illustration, in Fig. \ref{fig:shrinkageplot} we plot the posterior expectation ${\rm E}(\mu_{i}|\tau,\bm{D})$ 
of each of the best estimates of the distance moduli from each method as a function of
$\tau$ as $13$ coloured lines.
The black solid line is the MAP estimate $\hat{\gamma}(\tau)$ plotted as a function of
 $\tau$. When
 $\tau$ is close to zero, the conditional posterior 
means of each $\mu_{i}$
shrink toward the overall weighted mean $\hat{\gamma}(0)=\frac{\sum{D}_{i}/\sigma_{i}^{2}}{1/\sigma^{2}_{i}}$. The $\tau=0$ case corresponds to no systematic error and relatively large statistical
error.
 As $\tau$ becomes larger, the  conditional posterior means approach
 the case-by-case estimators of the distance moduli marked by plus signs at the far right in Fig. \ref{fig:shrinkageplot}.
 The red dashed vertical line indicates our estimate
 of $\tau$ and intersects the coloured curves at the hierarchical estimates of each $\mu_{i}$.
  Fig.~\ref{fig:shrinkageplot} shows how the hierarchical fit
 reduces to the case-by-case analyses as the variance of the systematic errors goes to infinity. We 
 include Fig.~\ref{fig:shrinkageplot} to
 illustrate the ``shrinkage'' of the estimates produced with hierarchical models, but such a plot
 is not needed to obtain the final fit.

\section{The Two-stage Algorithm for FB}\label{sec:fb}
In this section, we illustrate how to fit the hierarchical model (in Eq. \ref{eq:hierWD}) via
our two-stage algorithm. For more details about this algorithm, see \citet{si2017}.

\begin{description}
\item[\textbf{Step 0a:}] For each WD run BASE-9 to obtain a MCMC sample of ${p}(A_{i}, \bm{\Theta}_{i}|\bm{X}_{i})$ under the 
 case-by-case analysis. Thin each chain to obtain an essentially independent MCMC sample
 and label it $\{A_{1}^{(t)}, \bm{\Theta}_{1}^{(t)}, \cdots, A_{n}^{(t)}, \bm{\Theta}_{n}^{(t)}, t=1, 2, \cdots, t_{MC}\}.$
\item[\textbf{Step 0b:}] Initialise each WD age at $\tilde{A}_{i}^{(1)}=A_{i}^{(1)}$ and the other parameters at
$\tilde{\bm{\Theta}}_{i}^{(1)}=\bm{\Theta}_{i}^{(1)}$.\\
For $s=1, 2, \cdots$, we run Step 1 and Step 2.
        \item[\textbf{Step 1:}] Sample $\tilde{\gamma}^{(s)}$ and $\tilde{\tau}^{(s)}$ from 
${p}(\gamma, \tau| \tilde{A}_{1}^{(s)}, \cdots, \tilde{A}_{n}^{(s)})$.
        \item[\textbf{Step 2:}] Randomly generate $n$ integers between $1$ and $t_{MC}$, and denote
         them $r_{1},\cdots, r_{n}$.
For each $i$, set $A_{i}^{\ast}=A_{i}^{(r_{i})}, \bm{\Theta}_{i}^{\ast}=\bm{\Theta}_{i}^{(r_{i})}$
as the new proposal and
set $\tilde{A}^{(s+1)}=A_{i}^{\ast}, \bm{\Theta}^{(s+1)}=\bm{\Theta}_{i}^{\ast}$ with probability
$\alpha=\min\{1, \frac{{p}(A_{i}^{\ast}|\tilde{\gamma}^{(s)}, \tilde{\tau}^{(s)})/
{p}(A_{i}^{\ast}|\mu_{A_{i}}, \sigma_{A_{i}})}
{{p}(\tilde{A}_{i}^{(s)}|\tilde{\gamma}^{(s)}, \tilde{\tau}^{(s)})/
{p}(\tilde{A}_{i}^{(s)}|\mu_{A_{i}}, \sigma_{A_{i}})}\}$.
Otherwise, set $\tilde{A}^{(s+1)}=\tilde{A}^{(s)}, \tilde{\bm{\Theta}}_{i}^{(s+1)}=\tilde{\bm{\Theta}}_{i}^{(s)}$.
\end{description}
Steps 1 and 2 are iterated until a sufficiently large MCMC sample is obtained.
 If a good sample from the case-by-case analysis
is available, this two-stage sampler only takes a few minutes to obtain a MCMC sample from the
FB posterior distribution for the hierarchical model in Eq. \ref{eq:hierWD}.

\section{MCEM-type Algorithm}\label{sec:mcem}

In this section, we present our algorithm to optimise population-level
parameters in \textbf{Step 1} of EB-type methods (EB, EB-log and EB-inv).
We employ a Monte Carlo Expectation Maximisation (MCEM) algorithm
 with importance sampling for our EB-type methods. MCEM is a Monte Carlo
 implementation of Expectation Maximisation (EM) algorithm.
See \citet{van2010cross} for more details
 on EM and MCEM, and an illustration of their application
in astrophysics. To apply EM,
 we treat the object-level parameters, namely, 
$A_{1}, M_{1}, D_{1}, T_{1}, \cdots, A_{n}, M_{n}, D_{n}, T_{n}$ as latent variables. Due to
the complex structure of this astrophysical model, it is impossible to obtain
the expectation step (E-step) of the ordinary EM algorithm in closed form.  
MCEM avoids this via a Monte Carlo approximation to the E-step.
We employ two algorithms to compute the MAP estimate of $(\gamma, \tau)$: Approach 1 is MCEM and
Approach 2 uses importance sampling to evaluate the integral in
the expectation step instead of drawing samples from the conditional density of the latent variables.

Using Approach 1 to update $\gamma$ and $\xi=\ln\tau$ requires invoking
 BASE-9 once for each WD at each iteration of MCEM. This
is computationally expensive and motivates Approach 2.  
We suggest interleaving Approach 1 and 2 to construct a more
 efficient algorithm for computing the MAP estimates of $\gamma$ and $\tau$.

\noindent\textbf{Approach 1: MCEM} \\
\begin{itemize}[leftmargin=0pt,itemindent=30pt,topsep=-10pt]
  \item[\textbf{Step 0:}] Initialise $\gamma=\gamma^{(1)},\xi=\xi^{(1)}$, $d_{1}=1$ and $\tau=\exp(\xi^{(1)})$;\\
   Repeat for $t=1, 2, \cdots,$ until an appropriate convergence criterion is satisfied.
  \item[\textbf{Step 1:}] For star $i=1,\cdots, n$, sample $A_{i}^{[s,t]}, \bm{\Theta}_{i}^{[s,t]}, s=1,\cdots, S_{t}$
  from their joint posterior distribution
  \begin{equation*}
  {p}(A_{i}, \bm{\Theta}_{i}|\bm{X}_{i},\gamma^{(t)},\tau^{(t)})\propto
 {p}(\bm{X}_{i}|A_{i}, \bm{\Theta}_{i}){p}(A_{i}|\gamma^{(t)}, \tau^{(t)}){p}(\bm{\Theta}_{i}),
 \end{equation*}
  where $S_{t}$ is the MCMC sample size at the $t$-th iteration and should be an increasing function of
  $t$ (we take $S_{t}=1000+500t$).
  \item[\textbf{Step 2:}] Set\begin{align*}
\gamma^{(t+1)}&=\frac{1}{S_{t}\cdot{n}}\sum_{i=1}^{I}\sum_{s=1}^{S_{t}}A_{i}^{[s,t]},\\
\xi^{(t+1)}&=\log\bigg(\frac{1}{S_{t}\cdot(I-1)}\sum_{i=1}^{n}\sum_{s=1}^{S_{t}}(A_{i}^{[s,t]}-\gamma^{(t+1)})^{2}\bigg)/2;\\
\tau^{(t+1)}&=\exp(\xi^{(t+1)});\\
\end{align*}
\end{itemize}
\textbf{Approach 2: EM with importance sampling}\\
  Suppose we have a sample at the $\ast$-th iteration, $(A_{i}^{[\ast,s]}, \bm{\Theta}_{i}^{[\ast,s]}), i=1,\cdots, n, s=1,\cdots, S_{\ast}$ from the joint posterior distribution 
  ${p}(A_{i}, \bm{\Theta}_{i}|\bm{X}_{i},\gamma^{\ast},\tau^{\ast})$ given
  $\gamma=\gamma^{\ast}, \tau=\tau^{\ast}$.
 Set
\begin{align*}
w_{i}^{[t,s]}&=\frac{\frac{\phi(A_{i}^{[\ast,s]}|\gamma^{(t)},\tau^{(t)})}
{\phi(A_{i}^{[\ast,s]}|\gamma^{\ast},\tau^{\ast})}}
{\sum_{s=1}^{S_{t}}\frac{\phi(A_{i}^{[\ast,s]}|\gamma^{(t)},\tau^{(t)})}
{\phi(A_{i}^{[\ast,s]}|\gamma^{\ast},\tau^{\ast})}};\\
\gamma^{(t+1)}&=\frac{1}{n}\sum_{i=1}^{n}\sum_{s=1}^{S_{t}}A_{i}^{[\ast,s]}w_{i}^{[t,s]};\\
\xi^{(t+1)}&=\log\bigg(\frac{1}{(n-1)}\sum_{i=1}^{n}\sum_{s=1}^{S_{t}}\Large[A_{i}^{[\ast,s]}-
\gamma^{(t+1)}\Large]^{2}\bigg)/2;\\
\tau^{(t+1)}&=\exp(\xi^{(t+1)});\\
\end{align*}
where $\phi(x|\mu,\sigma)=\frac{1}{\sqrt{2\pi\sigma^{2}}}\exp(-\frac{(x-\mu)^{2}}{2\sigma^{2}})$.

\bsp	
\label{lastpage}
\end{document}